%% file: main.tex
\documentclass[
 reprint,
 superscriptaddress,
 nofootinbib,
 amsmath,amssymb,
 aps,prd
]{revtex4-2}

\usepackage[dvipsnames]{xcolor}
\usepackage{graphicx}
\usepackage{dcolumn}
\usepackage{bm}
\usepackage{placeins}
\usepackage{float}
\usepackage{comment}
\usepackage{graphicx}

\usepackage{hyperref}
\usepackage{makecell}

\usepackage{subfigure}

\usepackage{hhline}

\newcommand{\rom}[1]{\uppercase\expandafter{\romannumeral #1\relax}}

\input{journalheader}

\begin{document}

\preprint{}

\title{Exploring thermal effects of the hadron-quark matter transition in neutron star mergers}

\author{Sebastian Blacker}
\affiliation{Technische Universit\"at Darmstadt, Fachbereich Physik, Institut f\"ur Kernphysik, Schlossgartenstraße 9, 64289 Darmstadt, Germany}
\affiliation{GSI Helmholtzzentrum f\"ur Schwerionenforschung, Planckstra{\ss}e 1, 64291 Darmstadt, Germany}

\author{Andreas Bauswein}
\affiliation{GSI  Helmholtzzentrum  f\"ur  Schwerionenforschung,  Planckstra{\ss}e  1,  64291  Darmstadt,  Germany}
\affiliation{Helmholtz Research Academy Hesse for FAIR (HFHF), GSI Helmholtz Center for Heavy Ion Research, Campus Darmstadt,  Germany}

\author{Stefan Typel}
\affiliation{Technische Universit\"at Darmstadt, Fachbereich Physik, Institut f\"ur Kernphysik, Schlossgartenstraße 9, 64289 Darmstadt, Germany}
\affiliation{GSI Helmholtzzentrum f\"ur Schwerionenforschung, Planckstra{\ss}e 1, 64291 Darmstadt, Germany}

\date{\today}

\begin{abstract}
We study the importance of the thermal behavior of the hadron-quark phase transition in neutron star (NS) mergers. To this end, we devise a new scheme approximating thermal effects to supplement cold, barotropic equation of state (EoS) models, which is particularly designed for hybrid EoSs, i.e.~two-phase EoS constructions with a hadronic regime and a phase of deconfined quark matter. As in a previous, commonly adopted, approximate thermal treatment, we employ an ideal-gas component to model thermal pressure but, additionally, we include an improved description for the coexistence phase of hybrid EoSs. In contrast to the older scheme, our method considers the temperature dependence of the phase boundaries. This turns out to be critical for a quantitative description of quark matter effects in NS mergers, since the coexistence phase can introduce a strong softening of the EoS at finite temperature, which is even more significant than the change of the EoS by the phase transition at $T=0$. 
We validate our approach by comparing to existing fully temperature-dependent EoS models and find a very good quantitative agreement of postmerger gravitational-wave (GW) features as a figure of merit sensitively tracking the dynamics and thus the impact of quark matter in merger remnants. Simulations with the original thermal ideal-gas approach exhibit sizable differences compared to full hybrid models implying that its use in NS merger simulations with quark matter is problematic.
Importantly, our new scheme provides the means to isolate thermal effects of quark matter from the properties of the cold hybrid EoS and thus allows an assessment of the thermal behavior alone. 
Generally, we find that the thermal properties in hybrid models are more important compared to the thermal behavior of purely baryonic matter. We show that different shapes of the phase boundaries at finite temperature can have a large impact on the postmerger dynamics and GW signal for the same cold hybrid model. This finding demonstrates that postmerger GW emission contains important complementary information compared to properties extracted from cold stars in isolation or during a binary inspiral. We also show by concrete examples that it is even possible for quark matter to only occur and thus be detectable in finite-temperature systems like merger remnants but not in cold NSs. All these findings also illustrate that heavy-ion collision experiments as a probe of the phase diagram at finite temperature bear relevant information for the astrophysics of NS mergers and core-collapse supernovae. Furthermore, our new thermal treatment features the flexibility to be combined with any cold, barotropic hybrid model including effective models of phase transitions, where a large number of models is available. This allows to conduct large parameter studies to comprehensively understand the effects of quark matter in NS mergers. 

\end{abstract}

\maketitle
\section{Introduction}
An open question in neutron star (NS) physics is whether or not deconfined quark matter is present in the cores of NSs or in NS merger remnants, and a large number of studies have addressed this question in the last decades, e.g.~\cite{Glendenning1992,Heiselberg1993,Gentile1993,Steiner2000,Schertler2000,Gocke2001,Burgio2002,Banik2003,Baldo2003,Alford2003,Maieron2004,Buballa2004,Oechslin2004,Alford2005,Nakazato2008,Blaschke2009,Sagert2009,Weissenborn2011,Bonanno2012,Fukushima2013,Klaehn2013,Klaehn2017,Csaki2018,Drago2018,Paschalidis2018,Fischer2018,Blaschke2018,Christian2018,Most2018a,Annala2018,Burgio2018,Baym2018,Han2019,Sieniawska2019,Gieg2019,Orsaria2019,Chen2019,Alford2019,Dexheimer2019,Bauswein2019,Bauswein2019a,Montana2019,Ferreira2020,Ecker2020,Annala2020,Pang2020,Han2020,Chatziioannou2020,Drago2020,Blaschke2021,Pereira2021,Tang2021,Prakash2021,Jakobus2022,Bogdanov2022,Sumiyoshi2022,Huang2022,Pereira2022,Komoltsev2022,Bauswein2022,Tootle2022,Shirke2022,Pereira2022a,Zhu2022,Morawski2022,Gorda2022b,Takatsy2023,Annala2023}. At high baryon densities a phase transition from purely hadronic to deconfined quark matter is expected to occur eventually but it is not clear if this hadron-quark phase transition takes place at typical densities of NSs. Similar as the high-density equation of state (EoS) of nuclear matter, the regime of nonperturbative quantum chromodynamics (QCD) is challenging to describe theoretically. This is why even basic properties of the hadron-quark phase transition such as the onset density, the latent heat or the type of the phase transition, i.e.~its order, are not known~\cite{Haensel2007,Braun-Munzinger2009,Glendenning2012,Blaschke2018,Baym2018,Fukushima2011,Llanes-Estrada2019,Orsaria2019,Alford2019,Schaffner-Bielich2020}. 

For old, isolated NSs or NSs during the inspiral phase of a binary system temperature effects of the EoS can be neglected because they are too weak to affect the stellar structure. There exist a number of microphysical EoS models which include a phase transition but are limited to zero temperature (e.g.~\cite{Nambu1961,Farhi1984,Schertler2000,Gocke2001,Burgio2002,Alford2003,Banik2003,Maieron2004,Buballa2004,Alford2005,Kojo2015,Benic2015,Kaltenborn2017,Alford2017,Baym2019,Ivanytskyi2019,Alvarez-Castillo2019,Blaschke2020,Otto2020,Jokela2021,Xia2021,Kojo2022,Clevinger2022,Fraga2022,Blaschke2022}). EoSs with a hadron-quark phase transition are called hybrid EoSs since they join a model for a nuclear phase and a description for quark matter typically by employing a scheme to determine the phase transition such as the Maxwell construction~\cite{Glendenning1992,Glendenning2001,Hempel2009a,Hempel2013,Constantinou2023}. Also, several effective models have been put forward to incorporate a phase transition and a phase of quark matter in existing cold nuclear EoSs, i.e.~at zero temperature. These approaches are based for instance on piecewise polytropes~\cite{Alvarez-Castillo2017,Paschalidis2018,Weih2019,DePietri2019,Pang2020,Liebling2021,Pereira2022,Fujimoto2022,Kedia2022,Gorda2022b} or constant sound speed parametrizations of the quark phase~\cite{Alford2013,Christian2018a,Paschalidis2018,Christian2018,Burgio2018,Han2019,Montana2019,Han2020,Drago2020,Shahrbaf2021,Tang2021,Blaschke2021,Pereira2022a,Rau2022,Zhu2022,Guo2023}.

In core-collapse supernovae and NS mergers, temperature effects become relevant (see e.g.~\cite{Ruffert1997,Rosswog1999,Oechslin2007,Bauswein2010,Sekiguchi2011} and~\cite{Fischer2010,Huedepohl2010,Roberts2012}). Finite temperatures of several 10~MeV provide a sizable thermal pressure component and should thus be considered in numerical simulations and in the employed EoS. This also holds true for hybrid models implying that it is important to consider thermal effects when studying the impact of quark matter in such dynamical astrophysical systems.

On the one hand the inclusion of thermal effects introduces additional complexity for hybrid EoS models as they require to explicitly account for thermal effects in both phases of matter and to construct the phase transition for a whole range of temperatures assuming thermal, chemical and mechanical equilibrium. Usually this results in phase boundaries that vary with temperature and electrical charge fraction. Only a limited number of fully temperature-dependent hybrid EoS models exist to date (e.g.~\cite{Sagert2009,Dexheimer2010,PocahontasOlson2016,Klaehn2017,Fischer2018,Roark2018,Chesler2019,Cierniak2019,Malfatti2019,Motornenko2020,Bastian2021,Demircik2021,Prakash2021,Ivanytskyi2022}), and correspondingly few studies of NS mergers~\cite{Most2018a,Most2019,Bauswein2019,Bauswein2019a,Blacker2020,Bauswein2020,Bauswein2020a,Bauswein2021,Prakash2021,Tootle2022,Espino2023} or core core-collapse supernovae~\cite{Sagert2009,PocahontasOlson2016,Fischer2018,Fischer2021,Zha2021,Jakobus2022} with such EoS tables have been conducted.

On the other hand the temperature dependence of the EoS and of the phase boundaries is an inherent and characteristic property of the QCD phase diagram and therefore motivates a detailed investigation of these aspects~\cite{Fischer2014,Philipsen2019,Gao2020,Braun-Munzinger2022,Kumar2023}. For instance, the temperature dependence of the phase boundaries is critical to coherently connect the physics of NSs at zero or small temperatures and the insights from heavy-ion experiments, which probe the phase diagram at finite temperature of many 10~MeV and may reach further towards a critical endpoint~\cite{Masayuki1989,Andronic2006,deForcrand2008,Braun-Munzinger2022}.

Another instructive example is that most (if not all) existing temperature-dependent hybrid models for merger simulations show that at finite temperature the onset density of the hadron-quark phase transition occurs at lower densities compared to the onset density at $T=0$. This implies that in a NS merger reaching finite temperatures the sudden change of the EoS by the occurrence of quark matter would take place `earlier', i.e.~at lower densities, which may have a significant qualitative impact. These aspects exemplify the importance to understand not only the properties of the phase transition at $T=0$ but in particular the thermal behavior of the phase boundaries.

In this paper we assess the impact the temperature dependence of the hadron-quark phase boundaries can have on NS mergers. Importantly, we find that the detailed behavior can yield qualitatively different results with respect to the properties of the gravitational wave (GW) spectrum in the postmerger phase. This highlights the importance the behavior of the phase transition at finite temperature has and it demonstrates the value added by including (future) information from heavy-ion collision (such as~\cite{Hades2019,Friman2011,Senger2021,Blaschke2016,Abgaryan2022}), which can constrain the phase boundaries in this regime~\cite{Andronic2017,Dexheimer2021}.

Our exploration is largely based on an effective scheme describing thermal effects in hybrid EoSs, which we devise in this paper. This scheme, which adopts a commonly used thermal ideal-gas component~\cite{Janka1993}, allows to flexibly change the behavior of phase boundaries at finite temperature. Moreover, in the future it enables a straightforward incorporation of constraints on the phase boundaries from heavy-ion experiments.

As said, to date only a limited number of calculations exist for hybrid EoSs which provide the full temperature dependence, whereas numerous barotropic EoS models at $T=0$ are available. A number of hydrodynamical studies of NS mergers with hybrid EoSs thus employ an approximate treatment of the thermal pressure by adding the aforementioned ideal-gas component~\cite{Weih2019,DePietri2019,Ecker2020,Liebling2021,Fujimoto2022,Huang2022,Ujevic2022,Kedia2022,Guo2023}. For purely hadronic EoSs this ideal-gas component has been shown to yield quantitatively good results compared to a fully consistent treatment of temperature effects, e.g.,~\cite{Bauswein2010}. However, this approximation by construction cannot capture the temperature dependence of the phase boundaries, i.e.~of the occurrence of quark matter at finite temperature. In fact, we show that the common approximate thermal treatment does not qualitatively reproduce results of fully temperature-dependent hybrid EoSs. 

Our extension of the existing ideal-gas scheme to describe thermal effects in hybrid EoSs allows to model the temperature dependence of the phase boundaries such that we can quantitatively reproduce the results of existing temperature-dependent hybrid EoS models. In this work we only consider hybrid models which use a Maxwell construction of the phase transition.

We note that a major motivation to develop this new scheme is in fact to facilitate large-scale parameter studies with hybrid EoSs. Since thermal effects in hybrid models can be well captured by the new approximate thermal treatment, the large, currently already available variety of barotropic $T=0$ hybrid EoS models can be employed for numerical merger simulations. This includes also effective parametrizations of the phase transition and the quark matter EoS, which offers the advantage of a better coverage of the parameter space and to tune parameters and the properties of the resulting EoS in a controlled way. This may hardly be possible with full temperature-dependent hybrid EoS tables, which are more complex to build and more difficult to tune towards a desired behavior of the EoS using microscopic parameters of the model.

This paper is organized as follows: In Sect.~\ref{sec:scheme} we briefly discuss the commonly used thermal ideal-gas approach, demonstrate its shortcomings for hybrid EoSs and present an effective scheme to better capture the impact of temperature-dependent phase boundaries. We then use our new scheme to reconstruct a set of fully temperature-dependent hybrid EoS models in Sect.~\ref{sec:examples} and discuss parameter choices. In Sect.~\ref{sec:sims} we validate our approach in merger simulations and compare the results to models employing fully temperature-dependent EoSs and the traditional thermal ideal-gas approach. In Sect.~\ref{sec:applic} we present a discussion on how much the thermal behavior of hybrid EoSs influences the GW signal. For this we perform simulations with different phase boundaries at finite temperature together with the same cold EoS model. We summarize and conclude in Sect.~\ref{sec:sum}.

\section{Thermal effects in hybrid EoS}\label{sec:scheme}

As basis of our treatment and discussion of thermal effects in hybrid EoSs, we briefly summarize the commonly employed ideal-gas approach to approximately describe a thermal pressure component~\cite{Janka1993}. This scheme can be applied in combination with cold, barotropic EoSs and is thus often used in NS merger simulations since it allows to consider a much larger class of EoS models.

The ideal-gas treatment decomposes the pressure and the specific internal energy into a cold and a thermal part.

\begin{align}
    P &=P_\mathrm{cold}(\rho)+P_\mathrm{th} \label{eq:P}\\
    \epsilon &=\epsilon_\mathrm{cold}(\rho)+\epsilon_\mathrm{th} \label{eq:Eps}
\end{align}
$P_\mathrm{cold}$ and $\epsilon_\mathrm{cold}$ are functions of the density $\rho$ only and are determined by the barotropic model. $\rho$ and $\epsilon$ are evolved by solving the hydrodynamic equations. The thermal specific energy is then defined by $\epsilon_\mathrm{th}=\epsilon-\epsilon_\mathrm{cold}(\rho)$. 

From this the thermal part of the pressure $P_\mathrm{th}$ is obtained at a density $\rho$ adopting an ideal-gas like behavior
\begin{align}
    P_\mathrm{th}=(\Gamma_\mathrm{th}-1)\rho\epsilon_\mathrm{th}\label{eq:Pth}
\end{align}
where $\Gamma_\mathrm{th}$ is a chosen thermal ideal-gas index. In merger simulations, this value is usually assumed to be constant at all densities, even though microphysical EoS models show some variations~\cite{Constantinou2015}. For many models choosing $\Gamma_\mathrm{th}\approx1.75$ is a good compromise~\cite{Bauswein2010} and it is often chosen in merger simulations (see e.g.~\cite{Hotokezaka2013,Bernuzzi2015,Bauswein2019,Weih2019,Foucart2019,Ecker2020,Chaurasia2020,Blacker2020,Bauswein2020a,Bauswein2021,Liebling2021,Fujimoto2022,Palenzuela2022}.

Employing the ideal-gas scheme for thermal effects neglects the evolution of the electron fraction $Y_e$ and assumes that the composition effects are effectively captured. This has the practical advantage that any barotropic EoS model can be used, which does not need to provide the full dependence on the composition. A more elaborated scheme to approximate thermal effects which includes composition effects has been presented in~\cite{Raithel2019}.

We will show in the following that the usage of the thermal ideal-gas approach together with hybrid EoSs is problematic and can lead to qualitatively different results compared to fully temperature-dependent models.

\subsection{Phase boundaries of hybrid EoS}\label{subsec:phasebounds}
\begin{figure}
\includegraphics[width=1.0\linewidth]{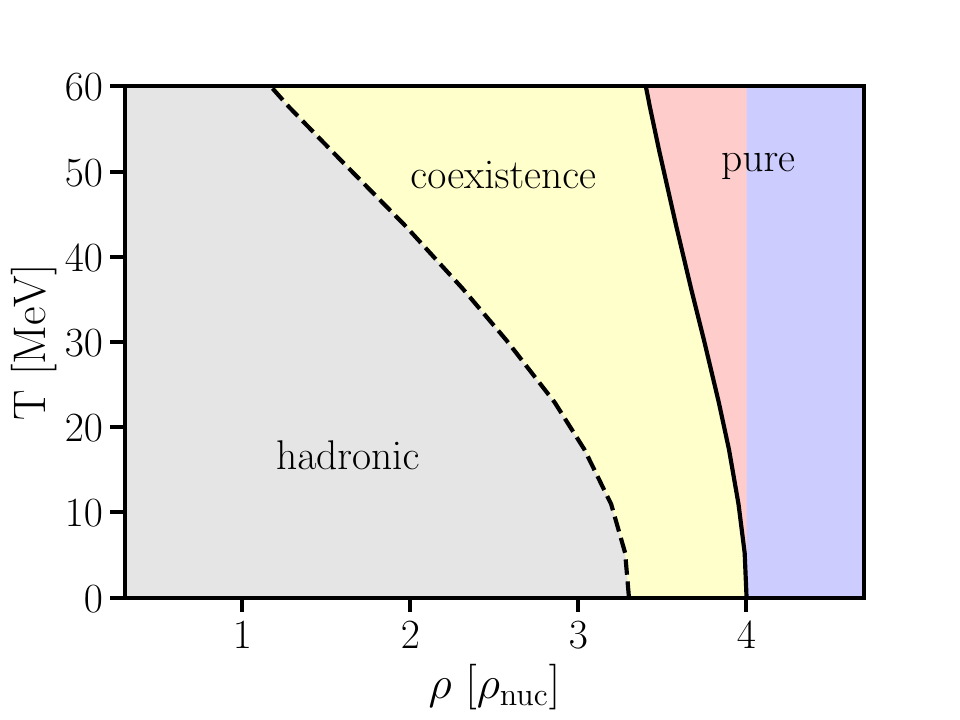}
\caption{Phase diagram of the DD2F-SF-1 EoS~\cite{Bastian2021} in the density-temperature plane. Different phases are highlighted with different colors. Black lines indicate the boundaries between these phases. The red colored region illustrates how much the pure quark phase is enlarged by the boundaries at finite temperatures. For this plot we pick the $Y_e$ values of the barotropic EoS, i.e.~those that correspond to cold, neutrinoless beta-equilibrium.}
\label{fig:DiagTRegimesComp}
\end{figure}

Even though the onset density and nature of the hadron-quark phase transition are currently unknown, it is plausible that the phase boundaries of this transition can vary significantly with temperature. As an example we show the phase diagram of the hybrid DD2F-SF-1 EoS~\cite{Bauswein2019,Bastian2021} as a function of temperature $T$ and rest-mass density $\rho$ in Fig.~\ref{fig:DiagTRegimesComp}. This $T$- and $Y_\mathrm{e}$-dependent model features a first-order phase transition from purely hadronic to deconfined quark matter including a region with coexisting hadron-quark phases. See Sect.~\ref{sec:Eosmodels} for more information on this EoS.

We highlight the purely hadronic and the coexistence phase region in gray and yellow, respectively. Additionally, we split the region of pure quark matter into a part where pure quarks are present at all densities (colored in blue) and a part where quarks only occur at finite temperatures (red area). The dashed line marks the onset of the coexisting hadron-quark phases and the solid line marks the transition to pure deconfined quark matter.

In this example the phase boundaries and thus the properties of the EoS significantly change with temperature, which is generally expected to be a feature of the QCD phase diagram.

Such a change in the EoS at finite temperature cannot be described within the simple ideal-gas approach, which approximates thermal effects in the same way at all densities and thermal energies. Therefore this treatment cannot properly capture finite-temperature effects in hybrid EoS models and is hence no longer a good approximation for the thermal pressure.

\begin{figure}
\includegraphics[width=1.0\linewidth]{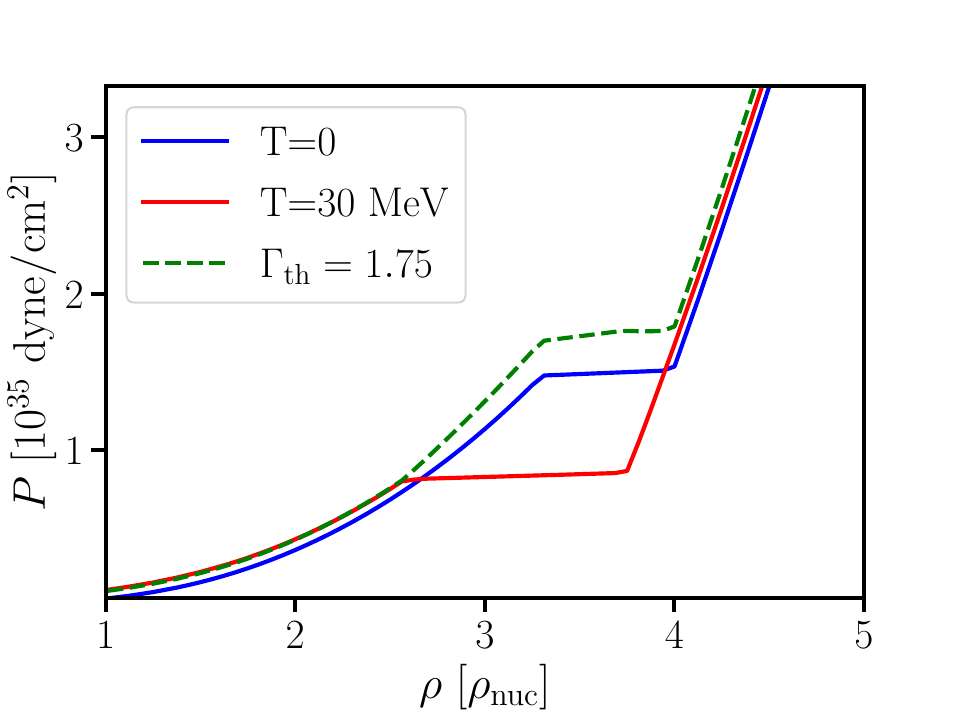}
\caption{Pressure as a function of density for the DD2F-SF-1 EoS~\cite{Bastian2021} at $T=0$ (blue line) and $T=30$~MeV (red line). The dashed green line indicates the pressure one would obtain by calculating the thermal pressure with the commonly used thermal ideal-gas approach of Eq.~\eqref{eq:Pth} with $\Gamma_\mathrm{th}=1.75$.}
\label{fig:EoSGammathProblem}
\end{figure}

We explicitly emphasize these issues in Fig.~\ref{fig:EoSGammathProblem}.
Here, the blue line shows the cold DD2F-SF-1 EoS in beta-equilibrium, i.e.~with an electron fraction such that $\mu_n=\mu_p+\mu_e$ at $T=0$ \footnote{We remark that the lowest tabulated temperature for the temperature-dependent EoS models we present here is $T=0.1$~MeV. For simplicity and clarity we adopt a slightly incorrect nomenclature and refer to this temperature as `$T=0$' or `cold' throughout this work.}, where $\mu_n$,$~\mu_p$,$~\mu_e$ are the neutron, proton and electron chemical potential, respectively.
The coexisting phases of hadronic and deconfined quark matter can be clearly seen as a flat region of almost constant pressure. At lower densities purely hadronic and at larger densities pure quark matter are present. This model employs a Maxwell construction to determine the transition between the two phases.

The red curve shows this EoS model at a temperature of 30~MeV and the same $Y_e$ profile across all densities. It is apparent that the transitions to the coexisting phases and the pure quark phase occur at lower densities compared to the $T=0$ case.
The flat region of the red curve is clearly shifted towards lower density values compared to the cold model. This shift is stronger for the onset of the coexisting phases than for the appearance of pure quark matter. Hence, the density range of the coexisting phases increases (see also Fig.~\ref{fig:DiagTRegimesComp}). In particular, the pressure in the coexisting phases is significantly lowered at finite temperature by nearly a factor of two implying a softening of the EoS.

The green, dashed curve shows the pressure one would obtain by using the ideal-gas approach with $\Gamma_\mathrm{th}=1.75$. Since the changes of the phase boundaries are not accounted for, the pressure $P_\mathrm{th}$ added by this treatment is always positive. Therefore, the total pressure is largely overestimated in the density range $\approx2.57\times \rho_\mathrm{nuc}$-$4.00\times \rho_\mathrm{nuc}$.

\begin{figure*}
\centering
\subfigure[]{\includegraphics[width=0.49\linewidth]{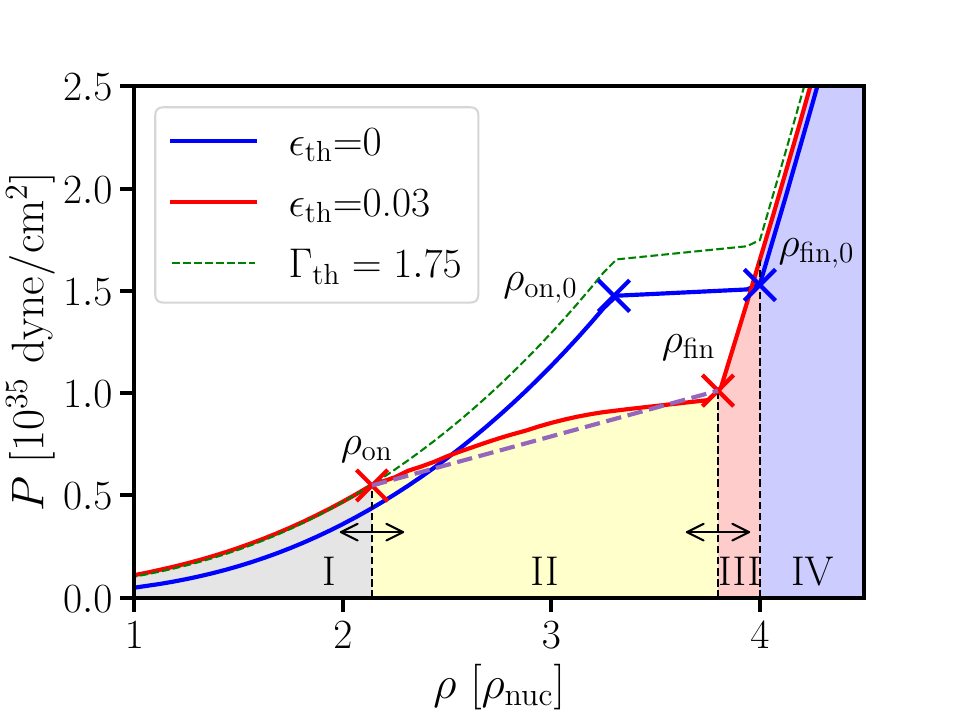}\label{fig:EoSRegimesComp1}}
\hfill
\subfigure[]{\includegraphics[width=0.49\linewidth]{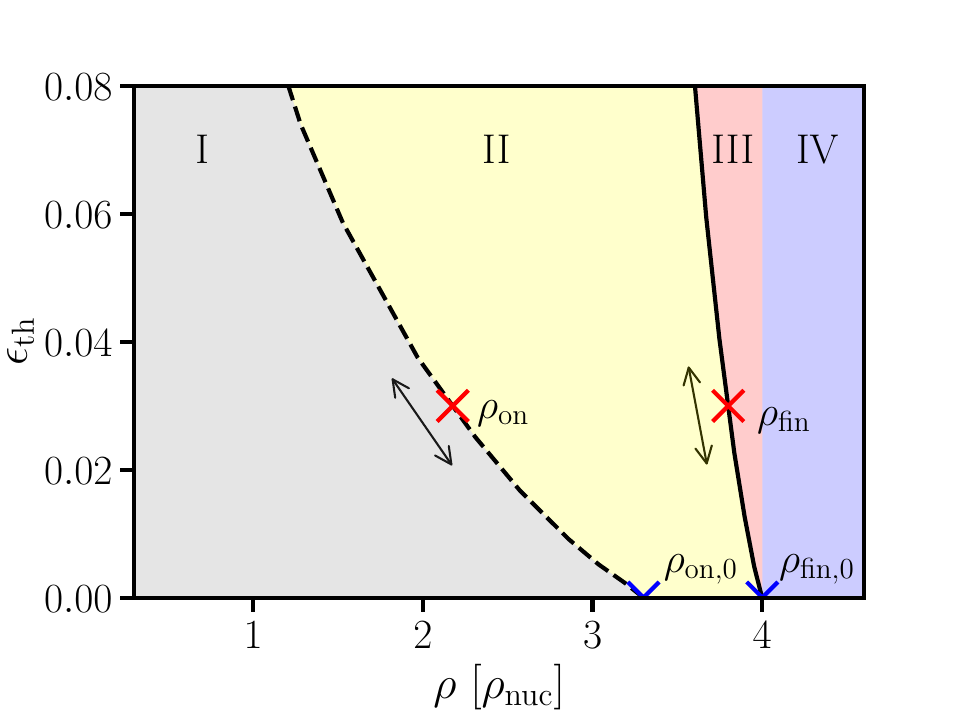}\label{fig:EoSRegimesComp2}}
\caption{(a): Pressure as a function of density for the DD2F-SF-1 EoS~\cite{Bastian2021} at $\epsilon_\mathrm{th}=0$ (blue line) and at $\epsilon_\mathrm{th}=0.03$ (red line). The green, dashed line displays the pressure one would obtain using the ideal-gas approach of Eq.~\eqref{eq:Pth} for $\epsilon_\mathrm{th}=0.03$. The purple, dashed line illustrates a linear interpolation between the phase boundaries at $\epsilon_\mathrm{th}=0.03$.\\
(b): Phase diagram of the DD2F-SF-1 EoS in the density-$\epsilon_\mathrm{th}$ plane. In both panels we highlight the different regimes as in Fig.~\ref{fig:DiagTRegimesComp}, which require a different treatment by our effective procedure to include thermal effects. Blue and red crosses refer to the phase boundaries at $\epsilon_\mathrm{th}=0$ and $\epsilon_\mathrm{th}=0.03$, respectively. Arrows indicate how these boundaries change with $\epsilon_\mathrm{th}$.}
\label{fig:EoSRegimesComp}
\end{figure*}
We emphasize that the transition densities and their shifts at finite temperature are very uncertain and modeldependent. The DD2F-SF-1 model we present here only serves as an example to highlight potential problems with the traditional ideal-gas approach when dealing with hybrid EoSs.

We also note that in quark matter $\Gamma_\mathrm{th}$ is typically close to 4/3 and thus a constant $\Gamma_\mathrm{th}$ may not provide an adequate description of all regimes in a hybrid model (see Sect.~\ref{sec:GammathChoice}).
Because of these problems we introduce a modified scheme to estimate the pressure at finite temperature better suited for these kinds of EoS models. We refer to this scheme as effective phase transition (effPT) scheme.

The main modification concerns the temperature-dependent phase boundaries and the resulting reduction of the pressure in the coexisting phases as well as a more realistic value of $\Gamma_\mathrm{th}=4/3$ in the quark phase.

\subsection{Effective thermal treatment for hybrid EoSs}\label{subsec:ourscheme}
As for the traditional approach (Eq.~\eqref{eq:Pth}) we use the specific thermal energy $\epsilon_\mathrm{th}$ instead of the temperature to determine the thermal pressure since this quantity is given by solving the hydrodynamic equations.

Accurate approximation of the pressure requires knowledge about how the phase boundaries depend on $\epsilon_\mathrm{th}$. The main idea is to estimate the pressure at the boundaries for finite $\epsilon_\mathrm{th}$ and interpolate between these values.

Working with $\epsilon_\mathrm{th}$ rather than $T$, in Fig.~\ref{fig:EoSRegimesComp1} we display the same DD2F-SF-1 EoS model as in Fig.~\ref{fig:EoSGammathProblem} but consider slices of constant $\epsilon_\mathrm{th}$. The blue line shows $\epsilon_\mathrm{th}=0$, which corresponds to $T=0$. This is the barotrope assumed to be available as a tabulated EoS within the hybrid approach. The red curve indicates the pressure of this microphysical model at a constant $\epsilon_\mathrm{th}=0.03$.

We label the boundaries of the coexisting phases for $\epsilon_\mathrm{th}=0.03$ with $\rho_\mathrm{on}$ and $\rho_\mathrm{fin}$ and for $\epsilon_\mathrm{th}=0$ with $\rho_\mathrm{on,0}$ and $\rho_\mathrm{fin,0}$ and highlight them with red and blue crosses, respectively. We emphasize that $\rho_\mathrm{on}$ and $\rho_\mathrm{fin}$ are functions of $\epsilon_\mathrm{th}$. The dashed purple line connects the phase boundaries at $\rho_\mathrm{on}$ and $\rho_\mathrm{fin}$. The pressure in the coexisting phases at $\epsilon_\mathrm{th}=0.03$ can be approximated by a linear relation. For our effPT scheme we will assume that the phase transition shows the typical features of a Maxwell construction, i.e.~two phases connected by a region of constant pressure at a given $T$ (see Fig.~\ref{fig:EoSGammathProblem}).

For comparisons we also display the pressure estimate from the traditional ideal-gas approach employing a constant $\Gamma_\mathrm{th}=1.75$ with a thin, dashed green line.

Once $\rho_\mathrm{on}$ and $\rho_\mathrm{fin}$ are known, we identify four different density regimes that each require a different treatment to estimate the pressure of hot matter from the cold, barotropic EoS.

The first regime corresponds to the densities below $\rho_\mathrm{on}$ at which both cold and hot matter with $\epsilon_\mathrm{th}=0.03$ are purely hadronic. The second regime covers the density range between $\rho_\mathrm{on}$ and $\rho_\mathrm{fin}$ in which hot matter has coexisting phases for the given $\epsilon_\mathrm{th}$. In the third regime between $\rho_\mathrm{fin}$ and $\rho_\mathrm{fin,0}$ hot material with $\epsilon_\mathrm{th}=0.03$ is in the pure quark phase while cold matter ($\epsilon_\mathrm{th}=0$) at the same density has coexisting phases. The fourth regime covers the density range above $\rho_\mathrm{fin,0}$ where pure deconfined quark matter is present at all temperatures. We highlight these four regimes in Fig.~\ref{fig:EoSRegimesComp} in gray, yellow, red and blue, respectively and label them with roman numerals \rom{1}-\rom{4}. The boundaries of the regimes are displayed by thin, vertical, dashed, black lines in Fig.~\ref{fig:EoSRegimesComp1}. 

The borders between regime \rom{1} and \rom{2} and between \rom{2} and \rom{3} will change depending on the value of $\epsilon_\mathrm{th}$. The border between \rom{3} and \rom{4} is fixed since $\rho_\mathrm{fin,0}$ is defined by the cold EoS.

To further illustrate these four regimes and how they change with $\epsilon_\mathrm{th}$ we plot the phase diagram of the DD2F-SF-1 EoS as a function of $\epsilon_\mathrm{th}$ and $\rho$ in Fig.~\ref{fig:EoSRegimesComp2}. We have highlighted the four regimes and the onset densities at $\epsilon_\mathrm{th}=0$ and $\epsilon_\mathrm{th}=0.03$ in the same way as in Fig.~\ref{fig:EoSRegimesComp1}, i.e.~with blue and red crosses. In Fig.~\ref{fig:EoSRegimesComp2} the dashed, black line marks the onset of the coexisting phases whereas the solid, black line displays the beginning of the pure quark matter regime. As indicated by the arrows different values of $\epsilon_\mathrm{th}$ will shift $\rho_\mathrm{on}$ and $\rho_\mathrm{fin}$ along the two phase boundaries.

\begin{figure}
\includegraphics[width=1.0\linewidth]{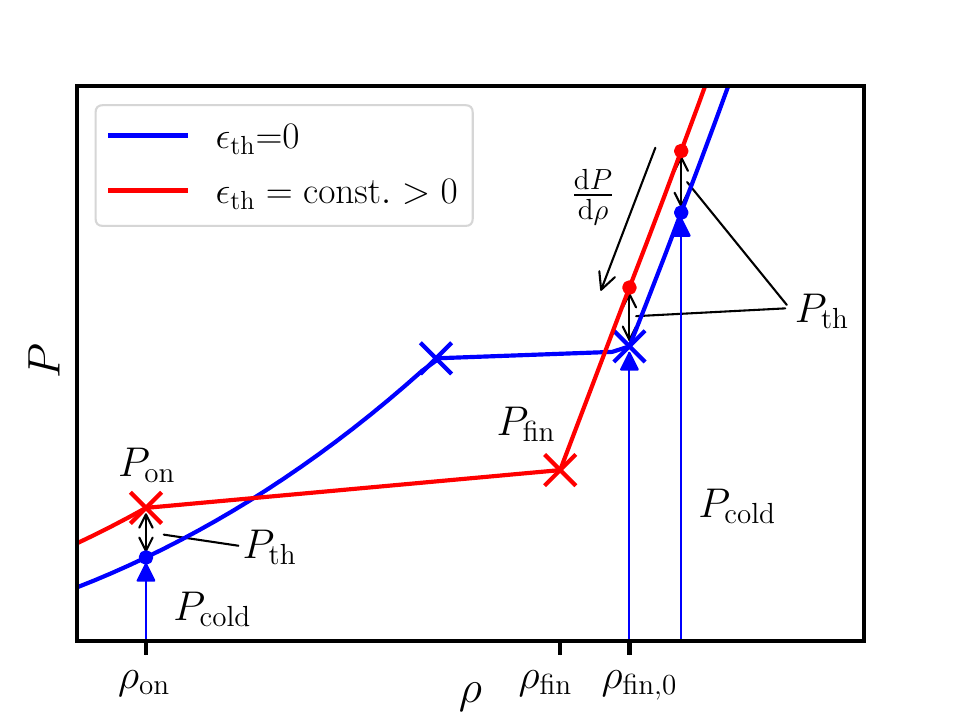}
\caption{Sketch of our effective procedure to estimate the pressure for $\epsilon_\mathrm{th}>0$. The blue line depicts a barotropic, hybrid EoS which we assume to be given. The red line represents this EoS at constant $\epsilon_\mathrm{th}>0$. For a given $\epsilon_\mathrm{th}$ $\rho_\mathrm{on}$ and $\rho_\mathrm{fin}$ are determined by the phase boundaries, which have to be provided. For the pressure $P_\mathrm{on}$ we estimate $P_\mathrm{th}$ from Eq.~\eqref{eq:Pth} with
$\Gamma_\mathrm{th}=1.75$. For $P_\mathrm{fin}$ we consider two points above $\rho_\mathrm{fin,0}$ and infer $P_\mathrm{th}$ at these points with Eq.~\eqref{eq:Pth} using $\Gamma_\mathrm{th}=4/3$. From these two points we obtain the slope $\frac{\mathrm{d}P}{\mathrm{d}\rho}$ and use it to extrapolate $P_\mathrm{fin}$ at $\rho_\mathrm{fin}$. For densities between $\rho_\mathrm{on}$ and $\rho_\mathrm{fin}$ we estimate $P$ through linear interpolation between $\rho_\mathrm{on}$ and $\rho_\mathrm{fin}$.}\label{fig:Sketch}
\end{figure}

Figure~\ref{fig:EoSRegimesComp2} visualizes the critical new input to our procedure to estimate thermal effects of hybrid EoSs: For given values of $\rho$ and $\epsilon_\mathrm{th}$, we determine in which of the four regimes I to IV matter with these thermodynamical properties is located. To this end we compare $\rho$ and $\epsilon_\mathrm{th}$ to the functions $\epsilon_\mathrm{th,on}(\rho)$ and $\epsilon_\mathrm{th,fin}(\rho)$. These represent the temperature-dependent phase boundaries between the purely hadronic and the region with coexisting phases and between the coexistence region and the pure quark matter state, respectively. Hence, in this approach we explicitly assume that $\epsilon_\mathrm{th,on}(\rho)$  and $\epsilon_\mathrm{th,fin}(\rho)$ are known functions. In practice, this can either mean assuming an explicit analytic function or providing tabulated values and using interpolation at nontabulated densities.

Note that for consistency the functions $\epsilon_\mathrm{th,on}(\rho)$ and $\epsilon_\mathrm{th,fin}(\rho)$ should reproduce the phase boundaries of the cold barotropic EoS, i.e.~$\rho_\mathrm{on,0}$ and $\rho_\mathrm{fin,0}$.

We now describe the specific treatments in the four different regimes and assume that in a first step $\rho_\mathrm{on}(\epsilon_\mathrm{th})$ and $\rho_\mathrm{fin}(\epsilon_\mathrm{th})$ have been determined for a given $\epsilon_\mathrm{th}$.

In the following subsubsections we will provide the derivation of the pressure at finite $\epsilon_\mathrm{th}$.

\subsubsection{Regime \rom{1} }\label{subsubsec:case1}

For $\epsilon_\mathrm{th}<\epsilon_\mathrm{th,on}(\rho)$ matter is in the purely hadronic regime. We therefore expect Eq.~\eqref{eq:Pth} to be a good approximation. From Fig.~\ref{fig:EoSRegimesComp1} one can see that this is indeed the case since the dashed, green and the red curve almost coincide at densities below $\rho_\mathrm{on}$. We hence use the ideal-gas approach with $\Gamma_\mathrm{th}=1.75$ in this regime.

\subsubsection{Regime \rom{2} }\label{subsubsec:case2}
For $\epsilon_\mathrm{th,on}(\rho) \leq \epsilon_\mathrm{th} < \epsilon_\mathrm{th,fin}(\rho)$ matter is in the coexistence region of Fig.~\ref{fig:EoSRegimesComp2}, i.e.~between $\rho_\mathrm{on}$ and $\rho_\mathrm{fin}$. At a given density the pressure at finite temperature may be below that of the cold barotropic EoS, and our treatment thus requires special care.

We outline the procedure in Fig.~\ref{fig:Sketch}. This figure shows a sketch of a hybrid EoS at $\epsilon_\mathrm{th}=0$ and at a constant, nonzero value of $\epsilon_\mathrm{th}$ with blue and red curves respectively, similar to Fig.~\ref{fig:EoSRegimesComp1}. Note that the lines in this figure are only meant to explain our procedure and do not show an actual EoS model.

The main idea is to determine the points $P_\mathrm{on}$ at $\rho_\mathrm{on}$ and $P_\mathrm{fin}$ at $\rho_\mathrm{fin}$ highlighted by red crosses in Fig.~\ref{fig:Sketch}. For this we calculate the densities $\rho_\mathrm{on}$ and $\rho_\mathrm{fin}$ by inverting the given phase boundaries. We then estimate $P_\mathrm{on}$ and $P_\mathrm{fin}$ at these densities, i.e.~at the edges of the coexisting phases.

For determining $P_\mathrm{on}$ we first read off the cold pressure $P_\mathrm{cold}$ at this density from the tabulated, cold EoS as indicated by the leftmost blue arrow in Fig.~\ref{fig:Sketch}. We then estimate the additional thermal pressure $P_\mathrm{th}$ using Eq.~\eqref{eq:Pth} with $\Gamma_\mathrm{th}=1.75$. 

This approach at the boundary to purely hadronic matter is consistent with the treatment of purely hadronic matter in regime \rom{1} and thus smoothly joins the two prescriptions in regimes \rom{1} and \rom{2}.

To obtain the pressure $P_\mathrm{fin}$ at the end of the coexistence phase we cannot use this approach since $\rho_\mathrm{fin}$ is smaller than the transition density $\rho_\mathrm{fin,0}$ of cold matter. Instead we extrapolate the EoS $P(\rho)$ at constant finite $\epsilon_\mathrm{th}$ from the density regime where pure quark matter occurs at $T=0$, i.e.~at densities just above $\rho_\mathrm{fin,0}$ down to the density $\rho_\mathrm{fin}$.

We do this by determining the pressure $P=P_\mathrm{cold}+P_\mathrm{th}$ at two densities slightly above $\rho_\mathrm{fin,0}$, picking two points from the tabulated cold EoS. These points are illustrated with two red dots in Fig.~\ref{fig:Sketch}. $P_\mathrm{cold}$ at these densities is directly given by the cold EoS as highlighted by the two rightmost blue arrows. For $P_\mathrm{th}$ we again use the ideal-gas approach. As discussed below, thermal effects in pure quark matter are well described by $\Gamma_\mathrm{th}=4/3$ as opposed to the higher value of $\Gamma_\mathrm{th}\approx1.75$ approximating purely hadronic matter. We therefore adopt $\Gamma_\mathrm{th}=4/3$ to quantify the thermal pressure $P_\mathrm{th}$ in this phase and to obtain the total pressure at the red dots.

We then employ the estimated pressure at the two points slightly above $\rho_\mathrm{fin,0}$, i.e.~the red dots, and extrapolate the pressure to the lower density at $\rho_\mathrm{fin}$.

In Fig.~\ref{fig:Sketch} this is indicated by the black arrow with the slope $\frac{\mathrm{d}P}{\mathrm{d}\rho}$ and provides $P_\mathrm{fin}$.

Now that we have estimated the pressure $P_\mathrm{on}$ and $P_\mathrm{fin}$ at the boundaries of the coexisting phases, we use linear interpolation between the two points to obtain the pressure at any density $\rho$ in the coexisting phases for the given $\epsilon_\mathrm{th}$. We expect a linear interpolation to be sufficiently precise for estimating the pressure in the coexisting phases at constant $\epsilon_\mathrm{th}$. We find that this is the case in Fig.~\ref{fig:EoSRegimesComp1} as the dashed, purple line approximates the red line well in the coexisting phases. 

In rare cases we found that the extrapolated value of $P_\mathrm{fin}$ is smaller than $P_\mathrm{on}$. To avoid an unphysical behavior of decreasing pressure with density, we use the slope of the cold EoS $m$ in the coexisting phases, i.e.~$m=(P_\mathrm{cold}(\rho_\mathrm{fin,0})-P_\mathrm{cold}(\rho_\mathrm{on,0}))/(\rho_\mathrm{fin,0}-\rho_\mathrm{on,0})$, to extrapolate linearly from $P_\mathrm{on}$ to the density $\rho$ in these cases. 

\subsubsection{Regime \rom{3} }\label{subsubsec:case3}
For $\epsilon_\mathrm{th,fin} \leq \epsilon_\mathrm{th}$ and $\rho<\rho_\mathrm{fin,0}$ matter is in the pure quark phase but for $T=0$ matter at this density would be in the coexisting phases. It is evident from Figs.~\ref{fig:EoSRegimesComp} and~\ref{fig:Sketch} that for some densities in this regime the pressure of hot matter can still be lower than the pressure of cold matter.

In the sketch in Fig.~\ref{fig:Sketch} regime \rom{3} basically spans from the red cross $\rho_\mathrm{fin}$ to the red dot at $\rho_\mathrm{fin,0}$ and we employ a procedure similar to the determination of $P_\mathrm{fin}$. We estimate the two red dots as in the previous case (regime \rom{2}). We then extrapolate linearly from these two points to the lower density $\rho$. For consistency we perform an additional check.  As before we determine the pressures at $\rho_\mathrm{on}$ and the slope of the cold EoSs in the coexisting phases $m=(P_\mathrm{cold}(\rho_\mathrm{fin,0})-P_\mathrm{cold}(\rho_\mathrm{on,0}))/(\rho_\mathrm{fin,0}-\rho_\mathrm{on,0})$. We then use $P_\mathrm{on}$ and $m$ to extrapolate linearly to $\rho$ obtaining a pressure $P^{*}$. To avoid unreasonably small pressure we pick the maximum of $P^{*}$ and $P$ as our approximated pressure in this regime.

\subsubsection{Regime \rom{4} }\label{subsubsec:case4}
For$\rho > \rho_\mathrm{fin,0}$ matter is in the pure quark phase and cold matter at the same density, too. We treat this regime using the ideal-gas approach as in regime \rom{1} and we do not need to consider any additional issues. Since we are in the pure quark regime we adopt a value of $\Gamma_\mathrm{th}=4/3$ (see also Fig.~\ref{fig:Gammath}).

Note that within our effPT scheme we have implicitly assumed that the phase boundaries at finite $\epsilon_\mathrm{th}$ are shifted towards lower densities, which may not necessarily be the case. The treatment outlined here may however be easily modified to describe such cases as well.

We emphasize that the description of the phase boundaries $\rho_\mathrm{on}(\epsilon_\mathrm{th})$ and $\rho_\mathrm{fin}(\epsilon_\mathrm{th})$ is an integral part of this effPT scheme and has to be provided as input in addition to the cold barotropic EoS. We also remark that we neglect any impact of the electron fraction $Y_e$ on the transition region from hadrons to deconfined quark matter.

\section{Description of phase boundaries and EoS examples}\label{sec:examples}

\subsection{Example hybrid EoS models}\label{sec:Eosmodels}

We now compare the results we obtain with our effPT scheme with actual fully temperature- and composition-dependent microphysical hybrid models. For this we use the set of seven different hybrid EoSs of Refs.~\cite{Bauswein2019,Blacker2020}. These are based on the DD2F-SF model of~\cite{Fischer2018,Bastian2018,Bastian2021}.
We follow the notation of Ref.~\cite{Bauswein2019} and label the individual EoSs with DD2F-SF-n with n $\in$\{1,2,3,4,5,6,7\}. All of these EoSs feature a strong first-order phase transition to deconfined quark matter with a region of coexisting hadron-quark phases.

The phase transition is constructed by fulfilling the Gibbs conditions for both electric and baryonic charges on the phase boundaries and imposing global charge neutrality in the coexisting hadron-quark phases~\cite{Bastian2021}.

Each of these hybrid EoS has different onset densities $\rho_\mathrm{on,0}$ and $\rho_\mathrm{fin,0}$ and features different stiffening of the quark phase. The hadronic phase is identical for all EoSs.

All DD2F-SF models include isospin and temperature dependence. In particular, this means that the phase boundaries between hadronic, coexistence and pure quark phase vary with both temperature and composition. Plots of phase diagrams for the different DD2F-SF-n EoSs in the density-temperature plane can be found in Ref.~\cite{Bastian2021}.

\begin{figure}
\includegraphics[width=1.0\linewidth]{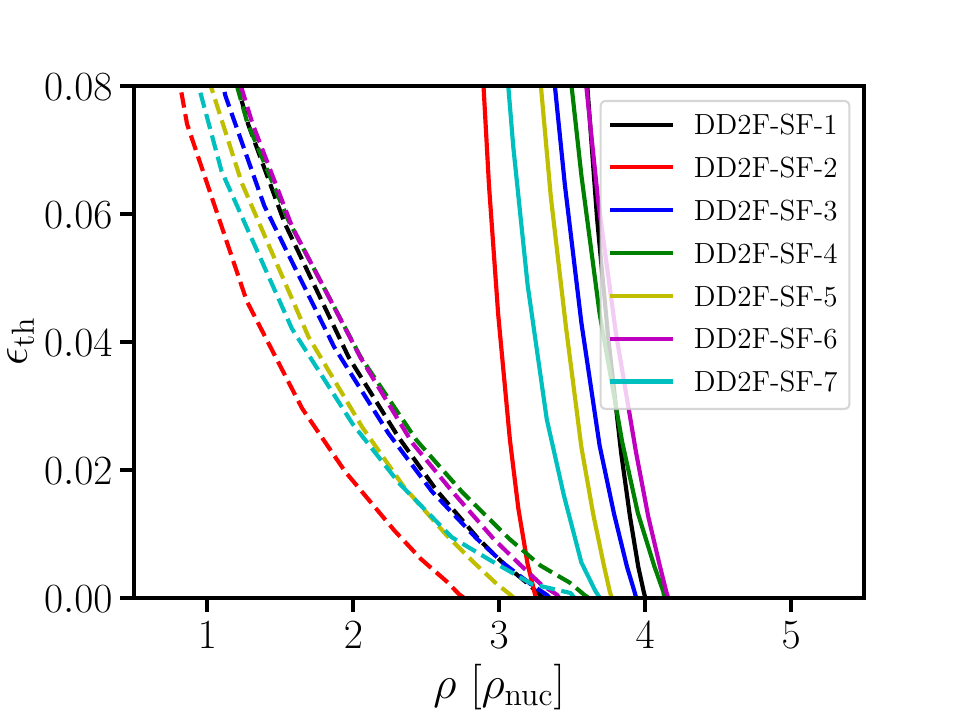}
\caption{Phase boundaries of the DD2F-SF-n EoSs~\cite{Bastian2021} in the density-$\epsilon_\mathrm{th}$ plane. Different colors refer to different EoS models. The dashed lines display the onset of the coexisting phases while the solid lines mark the beginning of the pure quark matter regime for each hybrid model.}
\label{fig:allphasebounds}
\end{figure}

In Fig.~\ref{fig:allphasebounds} we provide the phase boundaries of the hybrid DD2F-SF-n EoSs in the $\rho$-$\epsilon_\mathrm{th}$ plane. Different colors refer to different models. The dashed lines display the onset of the coexisting phases whereas the solid lines mark the beginning of the pure quark matter regimes. For each model we pick the same $Y_e$ profile at finite $\epsilon_\mathrm{th}$ across all densities as for the cold, beta-equilibrium case.

We see that the models differ in the onset densities, however, the overall behavior of the phase boundaries is quite similar for all EoSs. Generally all phase boundaries are shifted towards lower densities with increasing $\epsilon_\mathrm{th}$. This shift is larger for $\rho_\mathrm{on}$ than for $\rho_\mathrm{fin}$ meaning that the size of the coexistence phase region increases with larger $\epsilon_\mathrm{th}$.

We also find that EoSs with smaller $\rho_\mathrm{on,0}$ and $\rho_\mathrm{fin,0}$ also tend to have lower $\rho_\mathrm{on}$ and $\rho_\mathrm{fin}$ at finite $\epsilon_\mathrm{th}$ although crossing boundaries are possible.

In the following we refer to full temperature- and composition-dependent EoS tables as 3D tables and to the cold barotropic EoS in beta-equilibrium as 1D tables.

\subsection{Choices of $\Gamma_\mathrm{th}$}\label{sec:GammathChoice}

\begin{figure}
\includegraphics[width=1.0\linewidth]{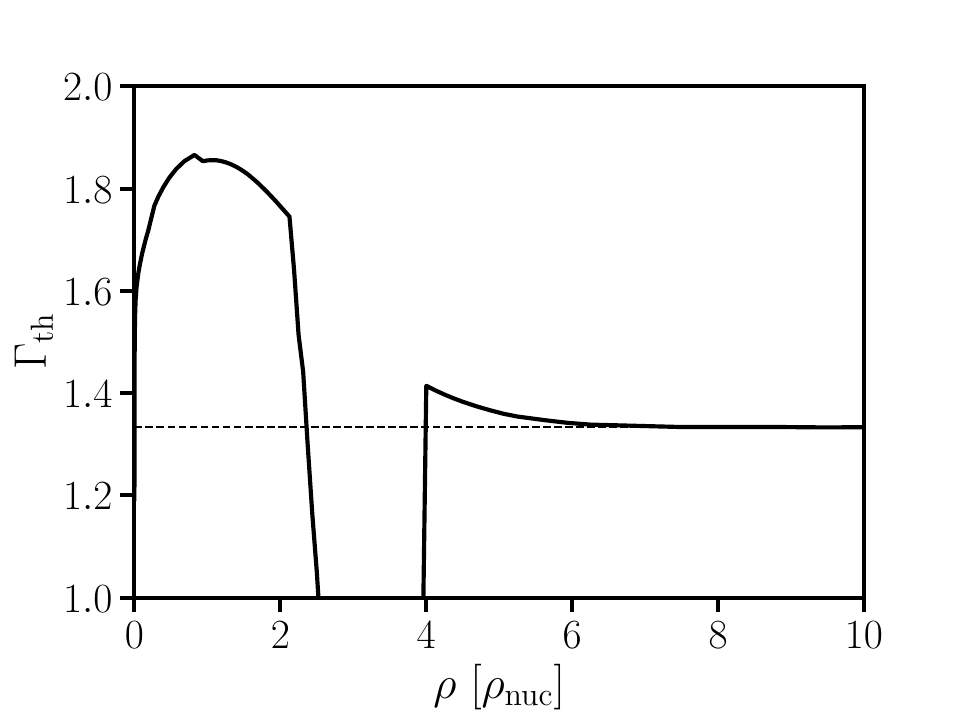}
\caption{Thermal ideal-gas index as function of density at $\epsilon_\mathrm{th}=0.03$ for the DD2F-SF-1 EoS~\cite{Bastian2021}. The dashed line marks $\Gamma_\mathrm{th}=4/3$, which is expected for an ultrarelativistic Fermi gas.}\label{fig:Gammath}
\end{figure}

We motivate the values of $\Gamma_\mathrm{th}$ which we have chosen in the different regimes of our effPT scheme. For this we determine the true pressure $P$ at $\epsilon_\mathrm{th}=0.03$ as a function of density from the 3D table of each DD2F-SF EoS. We then calculate the thermal pressure $P_\mathrm{th}$ using Eq.~\eqref{eq:P} where we infer $P_\mathrm{cold}$ from the 1D EoS table.
With this we can invert Eq.~\eqref{eq:Pth} to obtain $\Gamma_\mathrm{th}$ at different densities.

In Fig.~\ref{fig:Gammath} we show the inferred values of $\Gamma_\mathrm{th}$ at $\epsilon_\mathrm{th}=0.03$ as a function of density for the DD2F-SF-1 EoS. 

Note that for this model $\epsilon_\mathrm{eth}=0.03$ corresponds to temperatures of about 40~MeV to 30~MeV for densities between $2\times \rho_\mathrm{nuc}$ and $4\times \rho_\mathrm{nuc}$.

We find that at densities below $\approx 2\times \rho_\mathrm{nuc}$, where matter is in the hadronic phase, $\Gamma_\mathrm{th}$ is in the range of $\Gamma_\mathrm{th} \approx 1.6-1.85$. Thus, for regime \rom{1} a value of $\Gamma_\mathrm{th}=1.75$ is a sensible choice.

The following sharp drop of $\Gamma_\mathrm{th}$ is caused by the onset of the coexisting phases. Due to the `earlier' onset of the phase transition the pressure at finite $\epsilon_\mathrm{th}$ can be lower than $P_\mathrm{cold}$. Therefore, in this density range $P_\mathrm{th}$ as defined by Eq.~\eqref{eq:P} can formally be negative leading to $\Gamma_\mathrm{th}<1$ in this regime. 

In the pure quark phase $\Gamma_\mathrm{th}$ has a value of around 1.4 at the onset of the pure quark phase (at $\rho=4\times \rho_\mathrm{nuc}$ for the DD2F-SF-1 in Fig.~\ref{fig:Gammath}). With rising density it approaches 4/3, which is expected for an ultrarelativistic Fermi gas. We mark this value with a thin, dashed line in Fig.~\ref{fig:Gammath}.

Note that for simplicity in our effPT scheme we have chosen $\Gamma_\mathrm{th}=4/3$ at all densities in the pure quark phase including $\rho_\mathrm{fin,0}$ from which we extrapolate to lower densities. For this model we hence expect to slightly underestimate the pressure with our procedure in regime \rom{3} and at the low density part of regime \rom{4}. Also we expect to infer a slightly too low value of $P_\mathrm{fin}$ for the construction of the pressure in regime \rom{2}.  These errors are however not larger than those caused by choosing a constant $\Gamma_\mathrm{th}$ in the hadronic phase. 

We find similar results for the other DD2F-SF model and other values of $\epsilon_\mathrm{th}$ covering a wide range of temperatures. In particular, we find $\Gamma_\mathrm{th}\approx1.4$ around $\rho_\mathrm{fin,0}$ followed by a decay to $\Gamma_\mathrm{th}=4/3$ at larger densities.

We stress that the onset of quark deconfinement and the EoS of pure quark matter are very uncertain. It is hence not clear, whether this trend of $\Gamma_\mathrm{th}$ is a general behavior of hybrid EoS. Since all of the seven hybrid EoSs considered here use the same underlying SF model, it could also be a feature of these specific microphysical models.

\subsection{Direct comparison of EoS models} \label{sec:SchemeCompMean}
Before discussing simulations we demonstrate the validity of our scheme by reconstructing the DD2F-SF EoSs for a range of temperatures.

\begin{figure}
\includegraphics[width=1.0\linewidth]{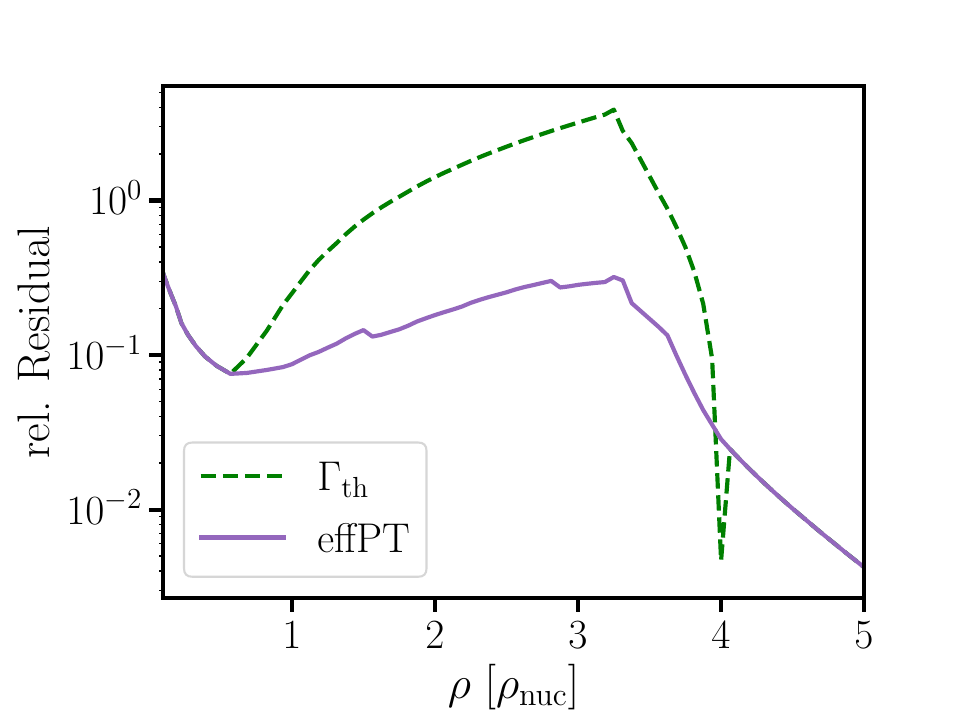}
\caption{Mean relative residuals of the total pressure from the thermal ideal-gas approach and our effective treatment of thermal effects in hybrid models compared to the actual DD2F-SF-1 EoS at different densities. At each density we average over the residuals in the temperature range between $T=5$~MeV and $T=80$~MeV.}
\label{fig:approachcomp}
\end{figure}

For this we consider the relative residuals of the pressure determined by our effPT scheme and by the commonly used ideal-gas approach with respect to the true pressure of the respective EoS. We define the relative residual as $|P_\mathrm{true}-P_\mathrm{approx}|/P_\mathrm{true}$, where $P_\mathrm{true}$ is the pressure inferred from the 3D EoS table and $P_\mathrm{approx}$ is the pressure approximated by either of the two schemes.

To at least somewhat account for the appearance of deconfined quarks in the traditional approach we now use different values of $\Gamma_\mathrm{th}$ at different densities. We pick $\Gamma_\mathrm{th}=1.75$ below $\rho_\mathrm{on,0}$, $\Gamma_\mathrm{th}=1.4$ between $\rho_\mathrm{on,0}$ and $\rho_\mathrm{fin,0}$ and $\Gamma_\mathrm{th}=4/3$ above $\rho_\mathrm{fin,0}$. Note that this choice of $\Gamma_\mathrm{th}$ leads to a drop in pressure at the densities $\rho_\mathrm{on,0}$ and $\rho_\mathrm{fin,0}$ for fixed $\epsilon_\mathrm{th}$.

To study the performance of both methods we average the relative residuals over several temperatures tabulated in the 3D EoS file of the DD2F-SF-1 model. We consider the range between $T=5$~MeV and $T=80$~MeV which is relevant for many astrophysical scenarios such as NS mergers or core collapse supernovae. In order to not over-represent low temperatures we keep a separation of at least 5~MeV between two considered temperatures.

In Fig.~\ref{fig:approachcomp} we plot the averaged residuals from our effPT scheme and from the traditional approach with a purple and a dashed green line, respectively. Note the logarithmic scale on the y-axis in this figure. 

We find that in the considered temperature range our effPT scheme is able to reproduce the true pressure of the DD2F-SF-1 model with much higher accuracy compared to the traditional approach at densities between $1\times \rho_\mathrm{nuc}$ and $4\times \rho_\mathrm{nuc}$. Especially in the density range between roughly $2\times \rho_\mathrm{nuc}$ and $3.5\times \rho_\mathrm{nuc}$ the average relative residuals of the effPT scheme are about an order of magnitude smaller compared to the traditional approach. This occurs even though we adjusted $\Gamma_\mathrm{th}$ as described above.

Note that in Fig.~\ref{fig:approachcomp} there appears to be a small region around $4\times \rho_\mathrm{nuc}$ where the traditional scheme seems to significantly outperform the effPT scheme. This behavior is caused by a single tabulated density point sitting right at the boundary between coexistence and pure quark phase. Here the true thermal pressure is very well approximated by $\Gamma_\mathrm{th}=1.4$ (compare Fig.~\ref{fig:Gammath}) whereas our effPT scheme uses $\Gamma_\mathrm{th}=4/3$ to extrapolate down from the pure quark phase.

We find similar results for the other DD2F-SF models. In order to further quantify our findings we average the relative temperature-averaged residuals over all tabulated densities of our EoS tables in the range between $0.5\times \rho_\mathrm{nuc}$ and $5\times \rho_\mathrm{nuc}$. We provide the values for all DD2F-SF EoSs in Tab.~\ref{tab:residuals} for both considered schemes.

\begin{table}
\begin{tabular}{c c c }
\hline\hline
EoS & traditional approach & effective scheme\\
\hline\
DD2F-SF-1 & 1.179 & 0.146 \\
DD2F-SF-2 & 1.429 & 0.149 \\
DD2F-SF-3 & 1.209 & 0.117 \\
DD2F-SF-4 & 1.111 & 0.089 \\
DD2F-SF-5 & 1.294 & 0.094 \\
DD2F-SF-6 & 1.139 & 0.090 \\
DD2F-SF-7 & 1.294 & 0.278 \\
\hline
\hline
\end{tabular}
\caption{Mean relative residuals of the total pressure calculated by the traditional ideal-gas approach and our effPT scheme compared to the actual EoS for all DD2F-SF models~\cite{Bastian2021} (compare Fig.~\ref{fig:approachcomp}). We average the residuals over all points tabulated in the respective EoS file in the temperature range of 5~MeV to 80~MeV and the density range of $0.5\times \rho_\mathrm{nuc}$ to $5\times \rho_\mathrm{nuc}$.}
\label{tab:residuals}
\end{table}
As in Fig.~\ref{fig:approachcomp} we find that our effective procedure performs very well overall and produces rather small average residuals of around 10\%-15\%. The only exception is the DD2F-SF-7 EoS where we find a somewhat larger average residual of about 28\%.

In the chosen density and temperature ranges the average residuals of the total pressure are generally roughly an order of magnitude smaller compared to the traditional ideal-gas approach for the considered sample of hybrid EoSs.

We conclude that at finite temperature it is extremely important to properly account for the shifting phase boundaries as they can potentially result in large pressure differences compared to models with fixed phase boundaries.

\section{Simulations and validation}\label{sec:sims}

\subsection{Setup}\label{sec:setup}
We further validate our effPT scheme by performing several NS merger simulations using the seven different hybrid DD2F-SF EoSs. For each model we simulate a merger with the 3D EoS table and a merger using only the 1D EoS table together with our effPT scheme. Additionally, we also perform a set of simulations using the 1D EoS tables together with the ideal-gas approach. To mimic the appearance of deconfined quarks in this approach we pick the same values of $\Gamma_\mathrm{th}$ as described earlier when discussing Fig.~\ref{fig:approachcomp}. 

To classify results from these different schemes we refer to simulations using the ideal-gas approach as $\Gamma_\mathrm{th}$ framework, simulations using the full 3D EoS table as 3D framework and simulations employing our effective phase transition scheme as effPT framework. We perform the simulations with a general relativistic, smoothed particle hydrodynamics (SPH) code~\cite{Oechslin2002,Oechslin2007} which solves the field equations using the conformal flatness condition~\cite{Isenberg1980,Wilson1996}.
A simulation starts from irrotational stars in cold, neutrinoless beta-equilibrium on a circular quasiequilibrium orbit with an initial separation of about 35~km (center to center). This configuration is relaxed for a short time with an artificial damping force to ensure an equilibrium SPH configuration at the beginning of the simulation. The stars then merge within a few orbits.

For the simulations within the 3D framework we assign the $Y_e$ value of beta-equilibrium to each SPH particle during the setup. Throughout the evolution of the system these values are then advected with the particles. For the other two frameworks the electron fraction of each SPH particle is always set to the value in beta-equilibrium according to its density as these frameworks neglect any potential impacts of $Y_e$.

For the models using the effPT framework we determine the phase boundaries, i.e.~$\rho_\mathrm{on}$ and $\rho_\mathrm{fin}$ of every DD2F-SF EoS as a function of the specific thermal energy $\epsilon_\mathrm{th}$ directly from the respective 3D table. At every density we interpolate to the $Y_\mathrm{e}$ corresponding to cold, beta-equilibrium. We then provide these boundaries in tabulated form to our code at the start of the simulation. 

\begin{figure}
\includegraphics[width=1.0\linewidth]{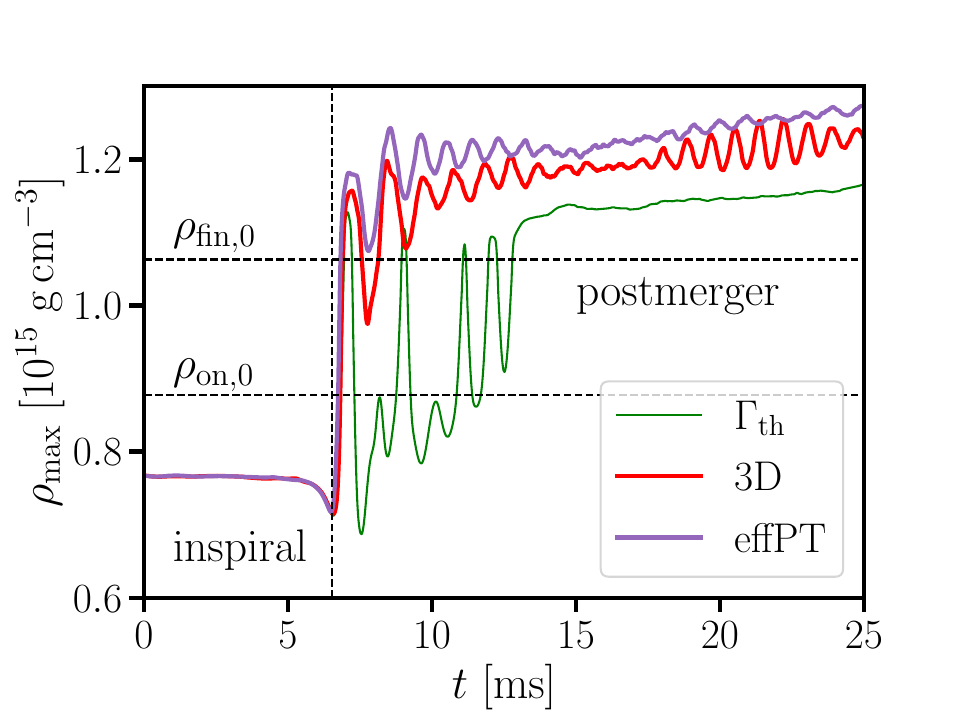}
\caption{Maximum rest-mass density as a function of time for the merger of two $1.35~M_\odot$ NSs using the DD2F-SF-1 EoS~\cite{Bastian2021} and the different frameworks outlined in the text to model the finite-temperature regime of the EoS. The horizontal dashed lines show the phase boundaries of the EoS at zero temperature.}
\label{fig:rhomaxs}
\end{figure}

\subsection{Simulation results}\label{sec:results}
In Fig.~\ref{fig:rhomaxs} we show the evolution of the maximum density during the merger of two $1.35~M_\odot$ stars with the DD2F-SF-1 EoS. We smoothed the simulation output of the maximum density throughout this paper since the SPH method features some small level of noise if quantities are directly evaluated on the particles. Additionally, we display the phase boundaries of this EoS model at zero temperature with dashed, horizontal lines.

Prior to the merger, the maximum densities are virtually identical in all three models because the stars are cold and the densities are below $\rho_\mathrm{on,0}=3.30\times \rho_\mathrm{nuc}$.

After the merger we find that matter at the maximum densities in the $\Gamma_\mathrm{th}$ framework bounces in and out of the quark phase as the merger remnant oscillates. Only at later times $\approx 6~\mathrm{ms}$ after the merger some material remains in the pure quark phase. 

The densities in the 3D framework do not show such a behavior. Here, matter at the center of the merger remnant enters the pure quark phase on the first contraction and remains in this state throughout the simulation. This is apparent from the overall larger densities and the smaller density oscillations compared to the $\Gamma_\mathrm{th}$ simulation.

The stronger density oscillations in the $\Gamma_\mathrm{th}$ framework are an artifact of the ideal-gas approach failing to properly account for the shifted phase boundaries with increasing temperature. From Fig.~\ref{fig:EoSRegimesComp1} we can infer that in addition to overestimating the pressure at certain densities this also leads to an incorrect EoS shape at finite $\epsilon_\mathrm{th}$. The changing phase boundaries smear out the transition region leading to a smoother pressure evolution with density whereas the ideal-gas approximation features a steeper increase of the pressure with density followed by a plateau of almost constant pressure.

Our effPT approach correctly considers the phase boundaries and hence is able to reproduce the maximum density evolution of the 3D framework simulation much more accurately. However, after the merger this scheme consistently slightly overestimates the densities in the merger remnant. This is in agreement with our previous findings (see Sect.~\ref{sec:GammathChoice}) that the effPT scheme slightly underestimates the thermal pressure in the regimes \rom{2} and \rom{3} leading to a more compact merger remnant.

Another possible source of error is that the effPT scheme does not capture isospin effects as matter is assumed to be in cold, beta-equilibrium composition at all times.

In three of our simulations (using the DD2F-SF-4,6,7 EoSs) employing the ideal-gas approach we observe a delayed transition occurring in the remnant. In these systems no quark matter is present after the initial density increase. With further evolution the remnant contracts and the transition sets in leading to a sudden increase of the maximum density several milliseconds after the merger.

We find that when using the 3D EoS tables or our effective approach this feature no longer occurs at the NS masses we consider here. Deconfined quark matter is always present right after the merger for all DD2F-SF EoSs. Hence, the delayed occurrence of deconfined quark matter in our simulated systems with the $\Gamma_\mathrm{th}$ framework is likely an artifact of neglecting temperature-dependent phase boundaries. This suggests that the delayed occurrence of quark matter in the early postmerger phase may be a less generic and common feature. We will further discuss this issue in Sect.~\ref{sec:otherEoSs}.

\begin{figure}
\includegraphics[width=1.0\linewidth]{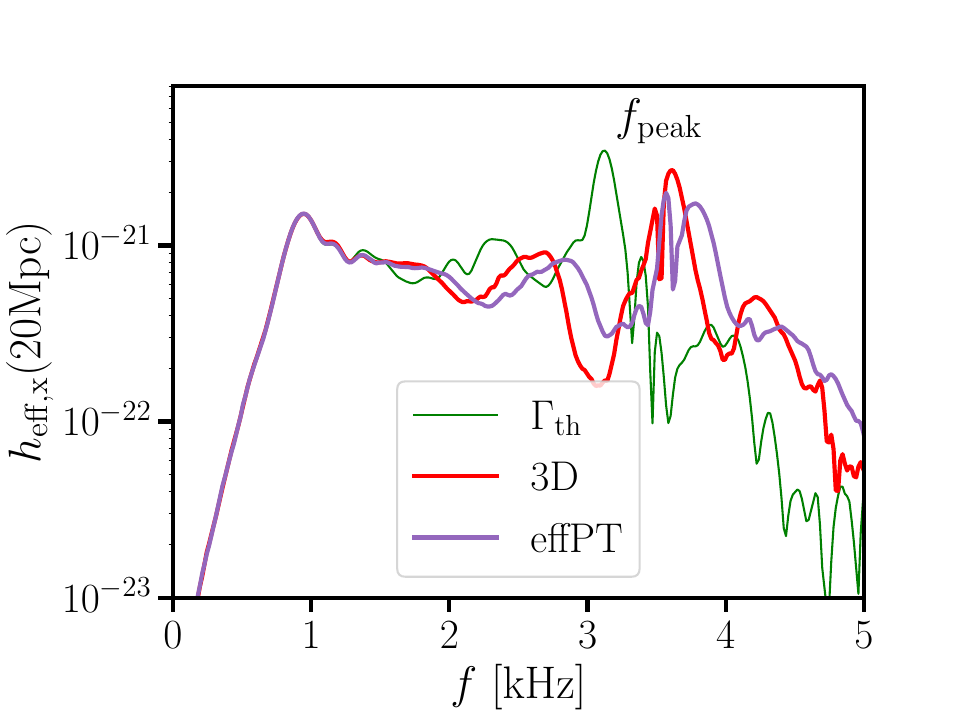}
\caption{Gravitational-wave spectrum of the cross polarization at a distance of 20~Mpc along the polar axis from the merger of two $1.35~M_\odot$ NSs using the DD2F-SF-1 EoS~\cite{Bastian2021}. Different colors represent results from the different frameworks to model the finite-temperature regime of the EoS.}
\label{fig:spectra}
\end{figure}

We now discuss the different GW signals produced by our simulated systems. Figure~\ref{fig:spectra} shows the spectra of the cross polarization at a distance of 20~Mpc along the polar axis for the same simulations as Fig.~\ref{fig:rhomaxs}. We use the same color scheme as in Fig.~\ref{fig:rhomaxs}.

The low frequency part of these spectra (roughly below 1.7~kHz) is formed during the inspiral of the two stars. The kHz range is produced by oscillations of the postmerger remnant (see~\cite{Chirenti2023} for an identification of distinct frequency peaks in the prompt emission of short gamma-ray bursts which match the frequencies found in simulated postmerger GW spectra). The dominant postmerger GW frequency $f_\mathrm{peak}$ characterizes the signal and scales with the remnant size~\cite{Bauswein2012}.

All three spectra agree well at low frequencies as the inspirals are mostly identical in our three models.

Using the $\Gamma_\mathrm{th}$ framework we infer an $f_\mathrm{peak}$ of 3.12~kHz whereas the value from the 3D framework is about 500~Hz larger with 3.61~kHz. As shown in Fig.~\ref{fig:EoSRegimesComp1} the $\Gamma_\mathrm{th}$ framework leads to a stiffer EoS and hence lower densities and a less compact remnant with smaller amounts of deconfined quark matter (see Fig.~\ref{fig:rhomaxs}).

The spectrum of the effPT framework in Fig.~\ref{fig:spectra} agrees much better with the results from the 3D framework. We deduce an $f_\mathrm{peak}$ value of 3.57~kHz for the effPT framework, which is about 40~kHz smaller than the value of the 3D framework. This is in slight tension with our earlier observation that the effPT framework tends to marginally overestimate the densities in the remnant.

A closer look at the spectrum of the effPT framework reveals that indeed the overall high frequency part does appear to be shifted a bit towards larger frequencies compared to the 3D framework. Only the maximum postmerger frequency $f_\mathrm{peak}$ is slightly smaller. However, there is a second peak in the spectrum of the effPT framework at higher frequencies (about 3.78~kHz). Such a split of the main peak is indicative of a drift in frequency. The frequency of the dominant mode changes with time as the structure of the merger remnant evolves. Hence the shape of the peak at $f_\mathrm{peak}$ is influenced by the dynamics of the merger remnant.

Generally, our effPT scheme is able to reproduce the overall shape of the GW spectrum and the value of $f_\mathrm{peak}$ with much higher accuracy than the ideal-gas approach. It also captures a secondary peak at about 2.5 kHz relatively well.

We provide the inferred $f_\mathrm{peak}$ values from the other DD2F-SF models in Tab.~\ref{tab:fpeaks}. Generally we find good agreement between the 3D framework and the effPT framework with a maximum difference in $f_\mathrm{peak}$ of 110~Hz. On the other hand we find larger deviations in $f_\mathrm{peak}$ when using the $\Gamma_\mathrm{th}$ framework with differences being as large as about 500~Hz. Interestingly, the frequencies we obtain within the $\Gamma_\mathrm{th}$ framework are relatively similar for all DD2F-SF models and also close to the $f_\mathrm{peak}$ value of 3.10~kHz (see~\cite{Bauswein2019,Blacker2020}) from the purely hadronic DD2F model~\cite{Typel2010,Alvarez-Castillo2016}. Within the ideal-gas approach the phase transition only has a minor impact on the overall structure of the merger remnant and hence the GW signal. It is understandable that the impact of deconfined quark matter is more significant for the 3D or the effPT scheme since the temperature dependence of the phase boundaries leads to the presence of deconfined quark matter already at lower densities, which consequently has a more significant influence on the overall remnant structure compared to the $\Gamma_\mathrm{th}$ simulations.

Especially we find that the occurrence of a delayed transition in the remnant some time after the merger does not leave a visible imprint in the GW spectrum. This is because the transition sets in at a time when the merger remnant has settled down and the GW emission is very weak. 

For the DD2F-SF-5 EoS we find that the spectra from the 3D and the effPT framework both exhibit two distinct dominant peaks at larger frequencies. The values from both frameworks agree within 60~Hz. Within the $\Gamma_\mathrm{th}$ framework this EoS also produces two distinct peaks in the high frequency part of the GW spectrum. However, the frequencies of these peaks disagree with peaks from the two other frameworks by about 300~Hz to 400~Hz. We provide the two values for each framework in Tab.~\ref{tab:fpeaks}.

\begin{table}
\begin{tabular}{c c c c }
\hline\hline
EoS & $\Gamma_\mathrm{th}$ & 3D & effPT\\
\hline\
DD2F-SF-1 & 3.12 & 3.61 & 3.57  \\
DD2F-SF-2 & 3.25 & 3.58 & 3.62 \\
DD2F-SF-3 & 3.12 & 3.50 & 3.52  \\
DD2F-SF-4 & 3.16 & 3.33 & 3.44  \\
DD2F-SF-5 & 3.22, 3.42 & 3.54, 3.81 & 3.60, 3.85 \\
DD2F-SF-6 & 3.14 & 3.64 & 3.67  \\
DD2F-SF-7 & 3.13 & 3.37 & 3.41  \\
\hline
\hline
\end{tabular}
\caption{Dominant postmerger GW frequency $f_\mathrm{peak}$ in kHz for our sample of DD2F-SF EoS models~\cite{Bastian2021} inferred from simulations. The different columns correspond to the different frameworks modeling the finite-temperature regimes as outlined in the text.}
\label{tab:fpeaks}
\end{table}

We conclude from these findings that it is mainly the shifting of the phase boundaries at nonzero temperatures that alters the GW signal enough to reveal the phase transition in the DD2F-SF models. Since such a shift cannot be described within the ideal-gas approach this method is incapable of capturing the effects of the transition to deconfined quark matter correctly. Therefore, it potentially greatly underestimates the postmerger GW frequencies. Our effPT scheme on the other hand does account for the changing phase boundaries at nonzero temperatures. It is hence able to reproduce results from the full 3D EoS table much more accurately. Especially we find that for the models we have tested the crude approximation of assuming that matter always has a composition as in cold, neutrinoless beta-equilibrium seems to be acceptable since the changes of the phase boundaries with $Y_e$ are small. The fact that the GW frequencies of the effPT scheme coincide well with the ones from the fully consistent simulations implies that it will similarly reproduce features that indicate the presence of quark matter in NS mergers such as the characteristic postmerger frequency shift relative to the tidal deformability as discussed in~\cite{Bauswein2019,Blacker2020}. In contrast, simulations with $\Gamma_\mathrm{th}$ lead to generally smaller frequencies compared to the fully temperature-dependent model and may thus not be able to reliably describe such features.

\subsection{Dependence of $f_\mathrm{peak}$ on the total binary mass}\label{sec:fpeakofmtot}
A general result from our previous work~\cite{Bauswein2019,Blacker2020,Bauswein2020a} was that $f_\mathrm{peak}$ is shifted to larger frequencies for the DD2F-SF models compared to the purely hadronic DD2F EoS. In~\cite{Blacker2020} we also discussed $f_\mathrm{peak}$ as a function of the total binary mass for the DD2F-SF models. We found that at low binary masses $f_\mathrm{peak}$ is practically identical for EoS models with and without a phase transition. In these systems the densities and temperatures are too low to trigger the deconfinement of enough matter to sufficiently alter the remnants structure necessary to substantially shift $f_\mathrm{peak}$.

With increasing mass the size of the quark core in the merger remnant grows and $f_\mathrm{peak}$ begins to deviate more and more from the purely hadronic value.

To provide a more stringent test we demonstrate that this behavior of $f_\mathrm{peak}$, i.e.~the quantitative deviance of $f_\mathrm{peak}$ in the purely hadronic and the hybrid models, is also correctly captured by our effPT scheme. For this we perform additional simulations using the DD2F-SF-6 and the DD2F EoSs. We simulate symmetric binaries with masses of $2.4~M_\odot$, $2.5~M_\odot$, $2.611~M_\odot$, $2.65~M_\odot$, $2.7~M_\odot$ and $2.78~M_\odot$, which are the masses we used in Ref.~\cite{Blacker2020}. For the DD2F-SF-6 EoS we perform simulations for every mass using the 3D and the effPT framework.

\begin{figure}
\includegraphics[width=1.0\linewidth]{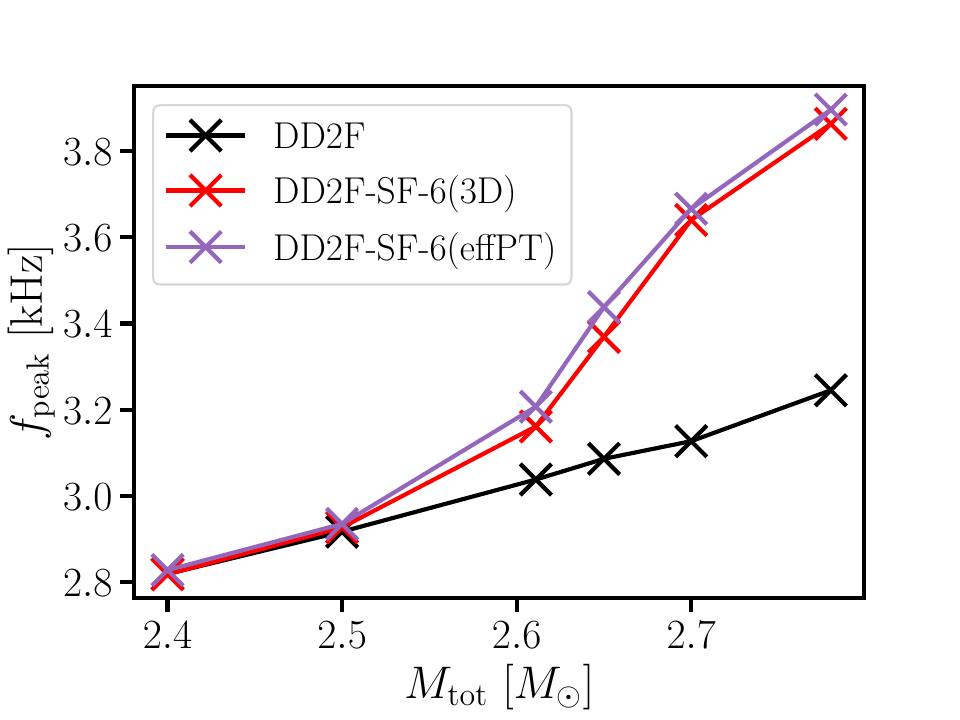}
\caption{Dominant postmerger GW frequency $f_\mathrm{peak}$ as a function of the total binary mass for the purely hadronic DD2F~\cite{Typel2010,Alvarez-Castillo2016} model (black) and the hybrid DD2F-SF-6 EoS~\cite{Bastian2021} (colored crosses). Crosses show simulation data, between these points linear interpolation is used. The two different colors refer to different approaches modeling the finite-temperature regime of the DD2F-SF-6 EoS as outlined in the text.}
\label{fig:fpeakMtot}
\end{figure}

We plot the inferred values of $f_\mathrm{peak}$ as a function of the total binary mass $M_\mathrm{tot}$ in Fig.~\ref{fig:fpeakMtot}.

We find that our effPT scheme correctly captures the behavior of $f_\mathrm{peak}$ as a function of $M_\mathrm{tot}$ and reproduces the frequencies of the 3D framework very well across the entire mass range. Especially the range in which the frequencies of the hybrid model start to deviate from the ones of the purely hadronic model agrees well.

As discussed before, we find that our effective procedure consistently slightly overestimates $f_\mathrm{peak}$ compared to simulations using the full temperature- and composition-dependent EoS tables. However, the differences we observe are small.

\section{Impact of temperature-dependent phase boundaries}\label{sec:applic}

In this section we apply the effPT scheme to explore the impact that different phase boundaries at finite $\epsilon_\mathrm{th}$ can have in merger simulations with a fixed cold EoS model. For this we consider four barotropic hybrid EoSs and perform simulations applying our effPT scheme with different assumed phase boundaries at finite $\epsilon_\mathrm{th}$ for the same cold EoS.

\subsection{Example 1: DD2F-SF based models}\label{sec:impact}
To explicitly probe finite-temperature effects we assume different phase boundaries at $T>0$ within our effPT framework together with the cold, beta-equlibrium composition slice of the DD2F-SF-7 model. We then compare the results to simulations using the actual boundaries. 

We remark that the phase boundaries of the DD2F-SF EoSs are constructed fulfilling the Gibbs condition for the pressure $P$ at constant temperature $T$ $P_1(\mu_b,\mu_q,T)=P_2(\mu_b,\mu_q,T)$, where $\mu_b$ and $\mu_q$ are the baryon and charge chemical potentials, respectively. It is obvious that this condition is only fulfilled at the true phase boundaries. 

However, we point out that our procedure describes the phase boundaries in terms of the specific thermal energy $\epsilon_\mathrm{th}$. The phase transition of the EoS on the other hand is constructed at constant temperature. Within different hadronic models there is considerable variation in $\epsilon_\mathrm{th}$ at a given density and temperature. See~\cite{Oertel2017} for a review on different EoSs models including finite temperatures.

Additionally, we stress that the knowledge of the EoS at zero temperature does not completely fix the EoS at finite temperature as different models can in principle lead to very similar cold EoSs but have considerable variation in the thermal part. In the case of hybrid EoS this can result in similar onset densities at $T=0$, and different phase boundaries at $T>0$. 

\begin{figure}
\centering
\includegraphics[width=1.0\linewidth]{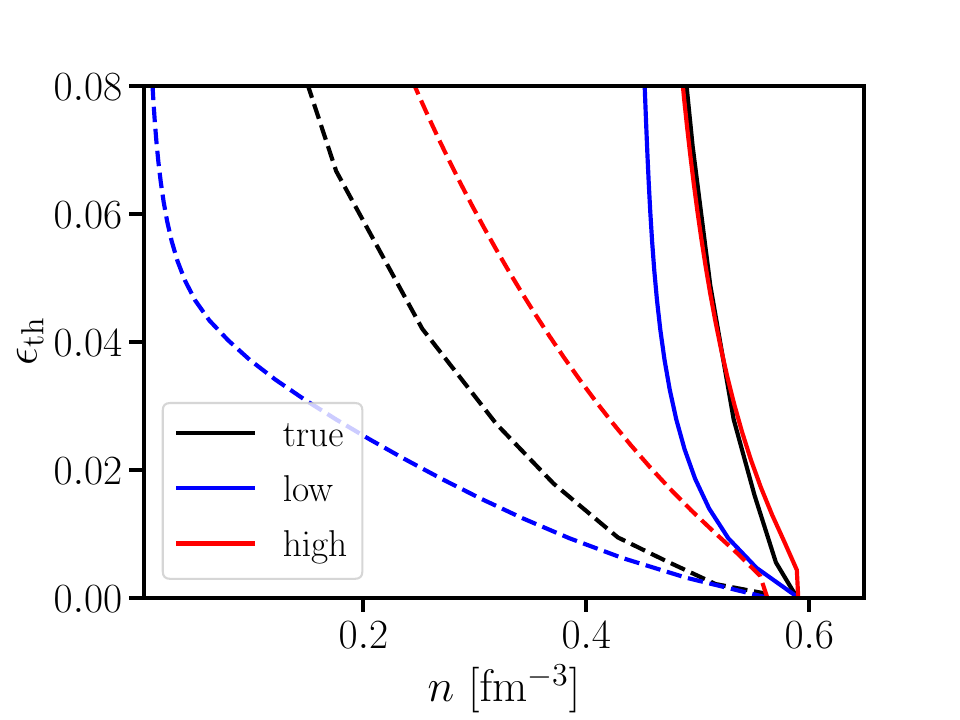}
\caption{Newly constructed phase boundaries in the $n$-$\epsilon_\mathrm{th}$-plane with our simple thermal EoS model (see App.~\ref{sec:shiftedbounds}) for the DD2F-SF-7 EoS~\cite{Bastian2021}. Different colors refer to different phase boundaries. Dashed lines mark the beginning of the coexisting phases and solid lines display the onset of pure deconfined quark matter. The nomenclature corresponds to the merger results shown in Fig.~\ref{fig:shifted}. Model true refers to the original phase boundaries of the DD2F-SF-7.}
\label{fig:NewBounds484}
\end{figure}

In App.~\ref{sec:TypelModel} we provide a simple, parametric approach to extend a barotropic EoS to finite temperature that obeys basic thermodynamic relations. For a given two-phase EoS at $T=0$ this allows to model the change of transition densities at finite temperature. We use this scheme to determine two alternative sets of phase boundaries for the DD2F-SF-7 EoS that coincide at $T=0$ but show different behaviors at $T>0$ while ensuring thermodynamic consistency.
We plot these new boundaries in Fig.~\ref{fig:NewBounds484} in the $\epsilon_\mathrm{th}-n$ plane with blue and red lines, respectively, where $n$ is the baryon number density. For a comparison we also plot the true boundaries of the DD2F-SF-7 EoS with black lines. Dashed (solid) lines mark the onset (end) of the coexisting phases.
In the following discussion we refer to the true phase boundaries as model true. From Fig.~\ref{fig:NewBounds484} we see one of our chosen parametrizations shifts the phase boundaries towards lower $\epsilon_\mathrm{th}$ compared to model true; we will refer to this model as model low. This mimics an `earlier' onset of the phase transition at finite temperature compared to the reference model. The second parametrization, which we will refer to as model high, on the other hand produces boundaries where the onset of quark deconfinement is moved to larger $\epsilon_\mathrm{th}$ corresponding to a `later' onset at finite temperature. The end of the coexistence phase is almost identical compared to the reference boundaries. As an example, at $\epsilon=0.04$ we find $\rho_\mathrm{on}=0.50\times \rho_\mathrm{nuc}$ and $\rho_\mathrm{fin}=2.92\times \rho_\mathrm{nuc}$ for model low, $\rho_\mathrm{on}=2.31\times \rho_\mathrm{nuc}$ and $\rho_\mathrm{fin}=3.25\times \rho_\mathrm{nuc}$ for model high and $\rho_\mathrm{on}=1.64\times \rho_\mathrm{nuc}$ and $\rho_\mathrm{fin}=3.25\times \rho_\mathrm{nuc}$ for model true.

To explore the impact of these changes of the EoS at finite temperature, we perform two additional simulations using the two newly constructed phase boundaries in our effPT scheme together with the cold DD2F-SF-7 EoS. We choose symmetric binaries with total system masses of $2.7~M_\odot$.

\begin{figure*}
\centering
\subfigure[]{\includegraphics[width=0.49\linewidth]{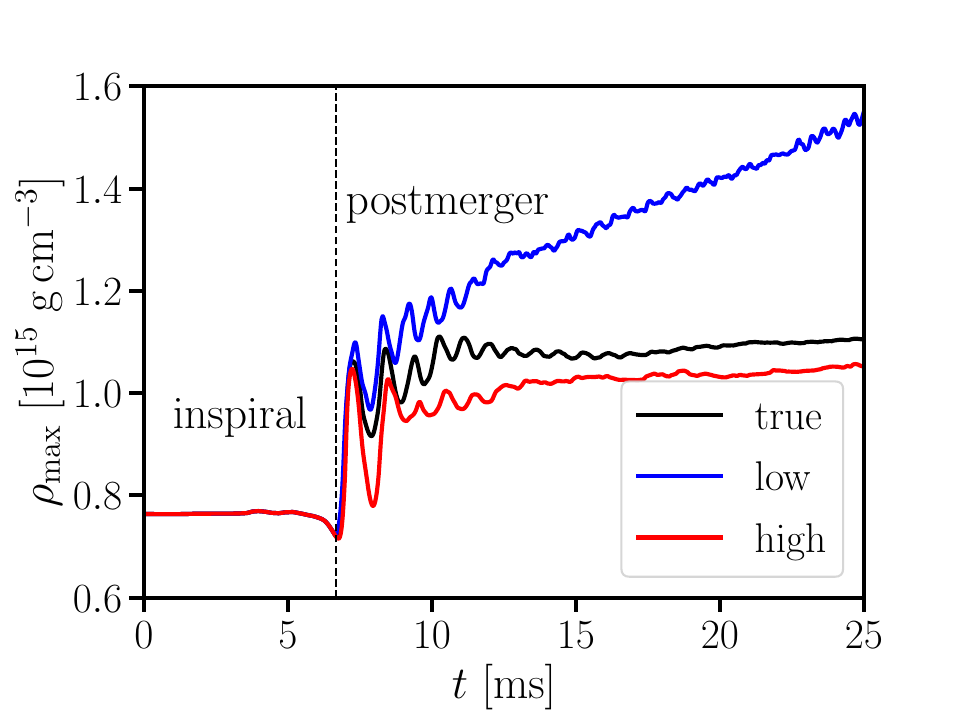}\label{fig:shifted_rhomax}}
\hfill
\subfigure[]{\includegraphics[width=0.49\linewidth]{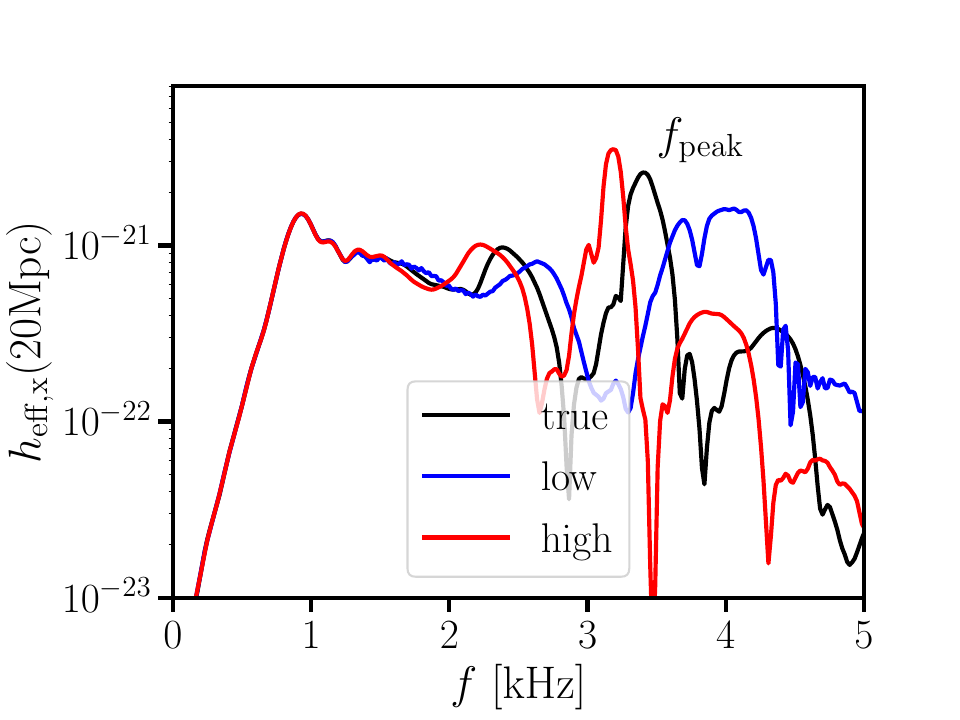}\label{fig:shifted_Spectra}}

\caption{(a): Evolution of the maximum rest-mass density for the merger of two $1.35~M_\odot$ NSs using the DD2F-SF-7 EoS~\cite{Bastian2021}. Different colors represent the different assumed shapes of the phase boundaries at finite temperature as shown in Fig.~\ref{fig:NewBounds484}. The EoS at zero temperature is identical for all three cases. (b): Gravitational-wave spectrum of the cross polarization at a distance of 20~Mpc along the polar axis from the same simulations as in (a).}
\label{fig:shifted}
\end{figure*}

In Fig.~\ref{fig:shifted} we compare the results from model true, model low and model high using black, blue and red lines, respectively.

Figure~\ref{fig:shifted_rhomax} shows the evolution of the maximum densities throughout the simulations. During the inspiral the densities are identical. After the merger they begin to deviate from each other. 

As expected we find that an `earlier' onset of the phase transition leads to larger postmerger densities with model low yielding the highest and model high resulting in the lowest density values.

Figure~\ref{fig:shifted_Spectra} shows the corresponding GW spectra of all three simulations. The postmerger signal is significantly affected by the shifted phase boundaries. Namely the value of $f_\mathrm{peak}$ is elevated by about 650~Hz for model low with respect to model true. In model high on the other hand we find that $f_\mathrm{peak}$ is about 220~Hz smaller compared to model true and comparable to results using the ideal-gas approach. These shifts are consistent with the behavior of the phase boundaries that lead to an `earlier'/`later' onset of the softening of the EoS by the appearance of deconfined quark matter.

This demonstrates that the behavior of the phase boundaries at finite temperature is of crucial importance for the diagnostics of the postmerger phase and the associated observables like $f_\mathrm{peak}$. The shape of the phase boundaries at finite temperature should thus be regarded as an important degree of freedom. We also refer to App.~\ref{sec:TypelModel}, which shows based on a simple model that the phase boundaries at nonzero temperatures are not fully determined by the cold EoS, exemplifying the crucial role of the postmerger phase to access this part of the QCD phase diagram.

We remark, that the finite-temperature effects, i.e.~the frequency shifts of the postmerger GW signal, are significantly more pronounced than those reported in other works for purely hadronic models, e.g.~\cite{Bauswein2010,Raithel2021,Fields2023}.

\subsection{Example 2: Piecewise polytropic models with low onset densities}\label{sec:otherEoSs}
As our effPT framework provides the flexibility to adopt chosen phase boundaries at finite temperature, we are in the position to apply this scheme to other EoS models employed in the literature. Hence, we can equip those models with a potentially more realistic behavior of the phase transition at finite temperature, i.e.~a phase boundary changing with temperature and bending more towards a critical point at lower densities and chemical potentials instead of a constant onset density and latent heat at all temperatures.

In Ref.~\cite{Weih2019} the authors used a piecewise polytrope model to represent an EoS with a soft coexistence phase starting at $2.085~\rho_\mathrm{nuc}$ and a stiff pure quark phase at densities above $4.072~\rho_\mathrm{nuc}$. They describe the hadronic phase below $2.085~\rho_\mathrm{nuc}$ with a piecewise polytropic representation of the relativistic mean field model FSU2H~\cite{Tolos2017,Tolos2017a}.

To capture thermal effects the authors employed the ideal-gas approach with $\Gamma_\mathrm{th}=1.75$. In a small mass range they observed a delayed onset of the phase transition shortly after the merger with a sudden increase in density. The authors observed that this was accompanied by a noticeable shift in the dominant postmerger GW frequency leading to two distinct peaks in the spectrum. Such a shift could hence serve as a clear indication of a first-order phase transition occurring in the remnant similar to Refs.~\cite{Bauswein2019,Blacker2020}.

Since we observe a comparable delayed transition in some of our DD2F-SF models when using the $\Gamma_\mathrm{th}$ framework, we perform three simulations with the EoSs model of Ref.~\cite{Weih2019}. We again choose symmetric binaries with total system masses of $2.7~M_\odot$. 

One simulation is conducted using the ideal-gas approach where we use the same values of $\Gamma_\mathrm{th}$ for the hadronic, the coexistence and the pure quark phase as in Sect.~\ref{sec:SchemeCompMean}. As before, we refer to this simulation as $\Gamma_\mathrm{th}$ simulation.

We perform two additional simulations using our effPT approach. Since this EoS only exists as a barotropic model, no finite-temperature phase boundaries are available. Furthermore, owing to the piecwise-polytropic parametrization the phase transition is not constructed in a consistent way, i.e.~matching pressure and chemical potentials at the transition densities. We find a difference of around 80~MeV in the chemical potentials at the borders of the coexistence phase. Because of the inconsistent chemical potentials, we cannot employ our model of App.~\ref{sec:TypelModel} to extend the phase boundaries to finite $\epsilon_\mathrm{th}$.

Instead we pick a functional form that somewhat resembles the phase boundaries of the DD2F-SF models. For the onset boundary we use a hyperbola
\begin{align}
    \epsilon_\mathrm{on}(\rho)&=\frac{1}{a_\mathrm{on}\rho+b_\mathrm{on}}+c_\mathrm{on}~~ \forall \rho <\rho_\mathrm{on} \label{eq:ethonWeih}
\end{align}
where $\rho$ refers to the rest-mass density in geometric units with $\mathrm{G}=\mathrm{c}=1$. We choose two sets of parameters with this phase boundary. For the first set we pick $a_\mathrm{on}=10065.12$, $b_\mathrm{on}=3.420353$ and $c_\mathrm{on}=-0.083468$, we refer to this simulation as effPTlow simulation. For the second set we chose $a_\mathrm{on}=4065.12$, $b_\mathrm{on}=1.420353$ and $c_\mathrm{on}=-0.205015$. This leads to a weaker shift of the phase boundaries towards lower densities as our first set, we refer to this simulation as effPThigh simulation. This choice of parameters reproduces the correct onset density of the cold EoS $\rho_\mathrm{on,0}=2.085 \times \rho_\mathrm{nuc}$ in both cases. For the end of the coexistence phase we have $\rho_\mathrm{fin,0}=4.072 \times \rho_\mathrm{nuc}$ at zero temperature. 

We recall that our effPT scheme uses a linear interpolation to estimate the pressure in the coexisting phases at finite $\epsilon_\mathrm{th}$ (see Sect.~\ref{subsubsec:case2}). We now construct the phase boundary of the pure deconfined quark matter phase by requiring at every $\epsilon_\mathrm{th}$ that slope of this linear interpolation is equal to the slope of the EoS in the coexisting phases at $T=0$, i.e.

\begin{align}
    \frac{P_\mathrm{fin}(\epsilon_\mathrm{th})-P_\mathrm{on}(\epsilon_\mathrm{th})}{\rho_\mathrm{fin}(\epsilon_\mathrm{th})-\rho_\mathrm{on}(\epsilon_\mathrm{th})}=\frac{P_\mathrm{fin,0}-P_\mathrm{on,0}}{\rho_\mathrm{fin,0}-\rho_\mathrm{on,0}}~~\forall \epsilon_\mathrm{th}~.
\end{align}
We chose this approach to ensure that for a fixed $\epsilon_\mathrm{th}$ the pressures inferred by the effPT scheme never decreases with density.

As an example with these parameters at $\epsilon_\mathrm{th}=0.03$ we infer $\rho_\mathrm{on}=1.24 \times \rho_\mathrm{nuc}$ and $\rho_\mathrm{fin}=3.51 \times \rho_\mathrm{nuc}$ for the effPTlow boundaries and $\rho_\mathrm{on}=1.62 \times \rho_\mathrm{nuc}$ and $\rho_\mathrm{fin}=3.63 \times \rho_\mathrm{nuc}$ for the effPThigh boundaries.

\begin{figure*}
\centering
\subfigure[]{\includegraphics[width=0.49\linewidth]{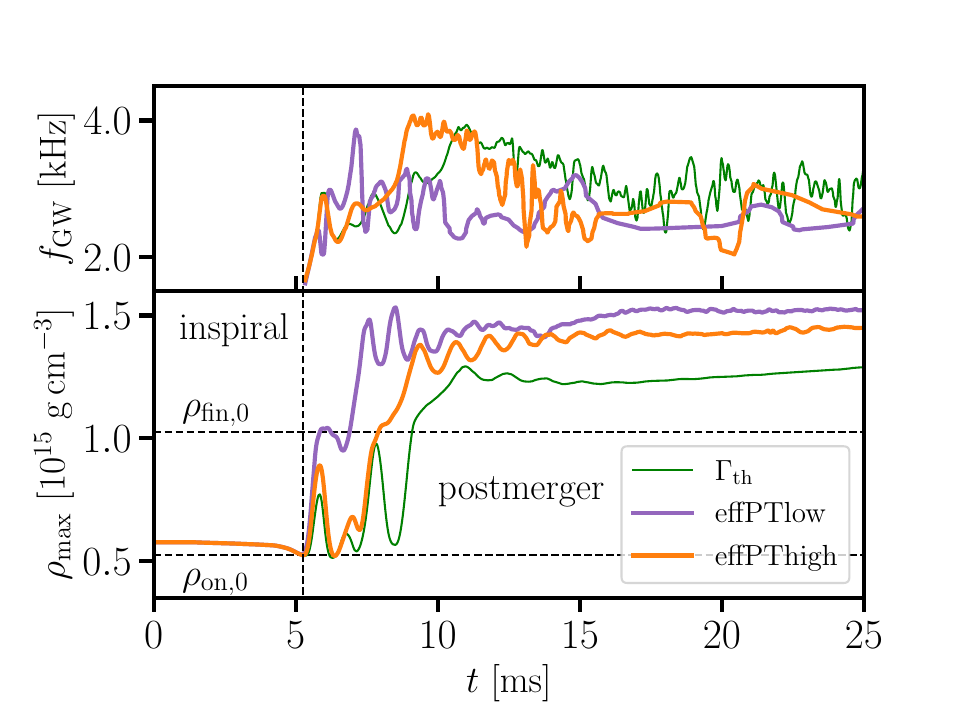}\label{fig:WeihppRhomax}}
\hfill
\subfigure[]{\includegraphics[width=0.49\linewidth]{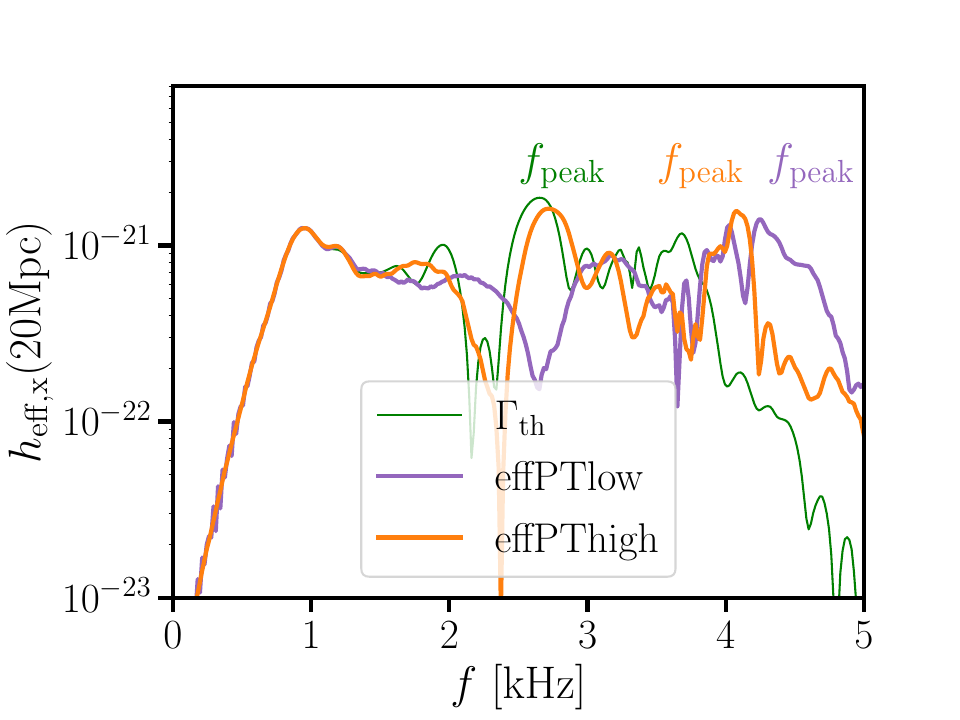}\label{fig:WeihppSpectra}}

\caption{(a): Top: Instantaneous gravitational wave frequency for the merger of two $1.35~M_\odot$ NSs using the piecewise polytropic EoS model of~\cite{Weih2019}. Different colors represent the different frameworks outlined in the text to model the finite-temperature EoS regime.
Bottom: Evolution of the maximum rest-mass density.
(b): Gravitational-wave spectrum of the cross polarization at a distance of 20~Mpc along the polar axis from the same simulations as in (a).}
\label{fig:Weihpp}
\end{figure*}

We compare the results from the $\Gamma_\mathrm{th}$ simulation, the effPTlow simulation and the effPThigh simulation in Fig.~\ref{fig:Weihpp} with green, purple and orange lines, respectively.

In the lower panel of Fig.~\ref{fig:WeihppRhomax} we plot the evolution of the maximum rest-mass density $\rho_\mathrm{max}$. Using the $\Gamma_\mathrm{th}$ framework we observe a similar trend of $\rho_\mathrm{max}$ as the authors of Ref.~\cite{Weih2019}. About three milliseconds after the initial bounce a delayed phase transition takes place in the merger remnant indicated by the strong increase in $\rho_\mathrm{max}$.

When using the effPThigh boundaries we see a similar behavior, however here the transition into the quark matter phase occurs on a shorter timescale and the maximum densities after the phase transition have increased.

If we employ our effPT procedure with the effPTlow phase boundaries we observe a different behavior. Now matter enters the pure quark phase directly on the second bounce after the merger. The maximum densities are also larger than in the two aforementioned frameworks. This is somewhat similar to our observations in Fig.~\ref{fig:rhomaxs}.

The differences in the postmerger remnant are also reflected in the GW spectra shown in Fig.~\ref{fig:WeihppSpectra}. The $\Gamma_\mathrm{th}$ simulation produces a pronounced peak at 2.65~kHz and additional peaks at larger frequencies up to 3.68~kHz. 

For the effPTlow simulation we find that the postmerger spectrum above about 1~kHz is greatly shifted towards larger frequencies compared to the results from the $\Gamma_\mathrm{th}$ simulation. Most notably the dominant postmerger frequency has a split peak at 4.24~kHz and 4.03~kHz. $f_\mathrm{peak}$ is about 1.6~kHz larger than the dominant frequency of the $\Gamma_\mathrm{th}$ simulation. This shift is much larger than the differences we observe between the $\Gamma_\mathrm{th}$ framework and the effPT framework when using the DD2F-SF EoSs and could be related to our specific choice of phase boundaries.

The effPThigh simulation yields two distinct peaks at 2.72~kHZ and 4.08~kHz. The lower frequency peak is comparable to $f_\mathrm{peak}$ of the $\Gamma_\mathrm{th}$ simulation whereas the higher frequency peak is close to the split peak structure of the effPTlow simulation. This behavior with two distinct well separated peaks in the GW spectrum was also reported in Ref.~\cite{Weih2019}. References~\cite{Hanauske2021,Hanauske2021a} noticed that the oscillations in $\rho_\mathrm{max}$ correlate with the instantaneous GW frequency. We calculate $f_\mathrm{GW}=\frac{1}{2\pi}\frac{\mathrm{d}(\phi(t))}{\mathrm{d}t}$ with $\phi(t)=\arctan(\mathrm{h}_\times(t)/\mathrm{h}_+(t))$, where $\mathrm{h}_\times$ and $\mathrm{h}_+$ are the cross and the plus polarized components of the GW signal. We plot the results for all three employed simulations in the upper panel of Fig.~\ref{fig:WeihppRhomax}. While we do find that some peaks of $f_\mathrm{GW}$ coincide with those of $\rho_\mathrm{max}$ and observe some similarities of the general trends, we generally do not see a too strongly correlated behavior of $f_\mathrm{GW}$ and $\rho_\mathrm{max}$ especially for the model with strongly temperature-dependent phase boundaries. We also note that Ref.~\cite{Hanauske2021a} showed that the exact behavior of $\rho_\mathrm{max}$ sensitively depends on numerical details.

Our examples demonstrate that important features in NS mergers with hybrid EoSs such as a delayed onset of a phase transition and the postmerger GW signal crucially depend on the behavior of the phase boundaries at finite $\epsilon_\mathrm{th}$. Simulations using hybrid EoSs models that do not include finite-temperature effects are hence potentially neglecting important physics that the simple ideal-gas approach cannot account for. 

\subsection{Example 3: Models with unstable hybrid branch}\label{sec:otherEoSs1}
\begin{figure}
\centering
\includegraphics[width=1.0\linewidth]{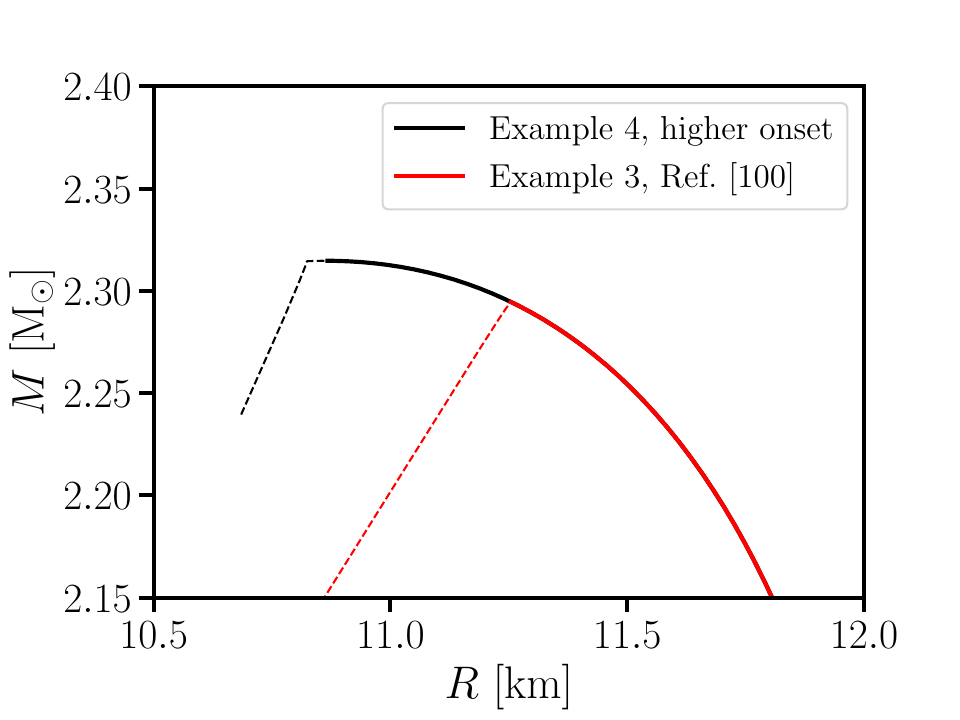}
\caption{Mass-radius relations for the cold EoSs used in Sect.~\ref{sec:otherEoSs1} and Sect.~\ref{sec:otherEoSs2}. Unstable stars are shown with thin dashed lines. Example~3 is the piecewise polytropic model of Ref.~\cite{Fujimoto2022} mimicking a strong first-order phase transition to deconfined quark matter that ends the mass-radius curve. Example 4 is a slightly modified EoS where the phase transition sets in just after the maximum mass is reached.}
\label{fig:FujimotoEoSs}
\end{figure}

The finite-temperature behavior of the phase boundaries can even lead to a qualitatively different outcome of the merger. We consider an EoS model of Ref.~\cite{Fujimoto2022}, which is based on a piecewise polytropic EoS with a density jump modeling a first-order phase transition. In this model the occurrence of quark matter yields a softening of the EoS such that no stable hybrid stars exist (see Fig.~\ref{fig:FujimotoEoSs}).

The onset density of quark deconfinement at $T=0$ is relatively high ($\rho_\mathrm{on,0}=5.36 \times \rho_\mathrm{nuc}$) corresponding to the central density of a 2.29~$M_\odot$ NS. In contrast to purely hadronic models and many hybrid EoS models, the mass-radius relation of stable nonrotating stars terminates at this point before reaching the typical high-mass regime of mass-radius relations where $\frac{d M}{d R}$ continuously approaches zero (see Fig.~\ref{fig:FujimotoEoSs}). In such a scenario no stable isolated nonrotating NSs with quark core can exist. Hence, inferring the presence of quark matter would require to precisely measure the properties of the mass-radius relation in this regime, which seems challenging.

To explore this scenario we perform two 1.4-1.4~$M_\odot$ simulations based on this cold EoS model. We set up one simulation with the $\Gamma_\mathrm{th}$ approach as in the original simulations in~\cite{Fujimoto2022} but with the same values for $\Gamma_\mathrm{th}$ as in Sect.~\ref{sec:SchemeCompMean} for the different phases. In this calculation the phase boundaries are thus assumed not to depend on temperature (as in the simulations of~\cite{Fujimoto2022}). In another simulation we include temperature-dependent phase boundaries with our effPT scheme. To construct these boundaries we use the same approach as for Example 2 with the parameters $a_\mathrm{on}=10065.12$, $b_\mathrm{on}=3.420353$ and $c_\mathrm{on}=-0.037540$ for Eq.~\eqref{eq:ethonWeih}. We then calculate the phase boundary of the pure deconfined quark matter phase using the same approach as outlined in Sect.~\ref{sec:otherEoSs} correctly reproducing $\rho_\mathrm{on,0}=5.36 \times \rho_\mathrm{nuc}$ and $\rho_\mathrm{fin,0}=8.91 \times \rho_\mathrm{nuc}$. This choice of parameters leads to a strong shift of the phase boundaries at finite $\epsilon_\mathrm{th}$. At $\epsilon_\mathrm{th}=0.03$ we obtain $\rho_\mathrm{on}=2.63\times \rho_\mathrm{nuc}$ and $\rho_\mathrm{fin}=3.57\times \rho_\mathrm{nuc}$.

These differences at finite temperature qualitatively affect the postmerger dynamics. In the $\Gamma_\mathrm{th}$ simulation, the system does not reach conditions for the formation of deconfined quark matter and the dominant GW postmerger frequency is $f_\mathrm{peak}=3.019$~kHz. Considering a temperature-dependent phase boundary in the effPT simulation, quark matter does occur after merging and the calculation yields $f_\mathrm{peak}=3.321$~kHz, which is higher than in the  $\Gamma_\mathrm{th}$ model since compared to the purely hadronic case the EoS softens due to the phase transition.

It is interesting that quark matter can apparently be present in  temporarily stable merger remnants but not in cold nonrotating NSs. We can trace back this behavior to the lower onset density at finite temperature noting that one can construct stable static NS solutions with quark core at finite temperature for the chosen treatment of phase boundaries within the effPT approach. We also perform simulations with higher total binary masses and do not find quark matter in metastable remnants within the $\Gamma_\mathrm{th}$ treatment. Investigating simulations with higher total binary mass for the $\Gamma_\mathrm{th}$ approach and the effPT scheme, we determine the threshold mass for prompt black-hole formation for both cases~\cite{Bauswein2021}. We find $M_\mathrm{thres}=3.07~M_\odot$ for the $\Gamma_\mathrm{th}$ case and $M_\mathrm{thres}=2.97~M_\odot$ within the effPT framework showing that the temperature dependence of the phase boundaries has an important effect on the outcome of NS mergers.

\subsection{Example 4: Models with high onset densities}\label{sec:otherEoSs2}
Another interesting example is models where the onset density of the hadron-quark phase transition at $T=0$ is larger than the maximum density of NSs (see Fig.~\ref{fig:FujimotoEoSs}). Hence, in such a situation observations of cold NSs could in principle not detect any signs of the quark deconfinement. However, if the onset density is considerably lowered at finite temperature, deconfined quark matter could occur in NS merger remnants and affect their evolution.

To explore this scenario we again adopt the model of Ref.~\cite{Fujimoto2022}. We slightly modify this EoS by shifting the onset density to $\rho_\mathrm{on,0}=6.22 \times \rho_\mathrm{nuc}$, which is higher than the central density of the most massive nonrotating NS (see Fig.~\ref{fig:FujimotoEoSs}). For the end of the coexistence phase we obtain $\rho_\mathrm{fin,0}=11.99 \times \rho_\mathrm{nuc}$. 

We perform two additional merger simulations based on this model. In the first simulation we use the traditional ideal-gas approach employing the same values of $\Gamma_\mathrm{th}$ as in Sect.~\ref{sec:SchemeCompMean} for the different phases. For this setup the phase boundaries do not depend on the temperature. In the second simulation we use our effPT scheme. We choose symmetric binaries with total masses of $2.8~M_\odot$.

We again construct the phase boundary for the effPT calculation as in Sect.~\ref{sec:otherEoSs}, where we pick $a_\mathrm{on}=10065.12$, $b_\mathrm{on}=3.420353$ and $c_\mathrm{on}=-0.032931$ for Eq.~\eqref{eq:ethonWeih}. With these parameters we obtain $\rho_\mathrm{on}=2.88\times \rho_\mathrm{nuc}$ and $\rho_\mathrm{fin}=4.91\times \rho_\mathrm{nuc}$ at $\epsilon_\mathrm{th}=0.03$.

From the $\Gamma_\mathrm{th}$ simulation we infer a dominant postmerger GW frequency of 3.019~Hz whereas the effPT simulation yields $f_\mathrm{peak}=3.192$~Hz. Note that the result from the $\Gamma_\mathrm{th}$ simulation is identical to Example 3 as no deconfined quark matter is present in these models and the hadronic EoS parts are identical. As for Example 3, in the effPT simulation the shift of the phase boundaries leads to an `earlier' appearance of deconfined quark matter, which significantly affects the dynamics of the postmerger remnant and its GW signal. 
An even stronger impact is found in simulations with slightly higher total binary masses of $2.86~M_\odot$, where we find a difference in the main postmerger frequency of 358~Hz ($f_\mathrm{peak}=3.059$~kHz for the $\Gamma_\mathrm{th}$ simulation; $f_\mathrm{peak}=3.417$~kHz for the effPT simulation). This shows that even if the hadron-quark phase transition takes place at densities not reached in cold NSs, it could still be accessible in NS mergers. This result again demonstrates the relevance of temperature effects of QCD phase diagram and the importance of postmerger GW 
emission.

\begin{figure}
\centering
\includegraphics[width=1.0\linewidth]{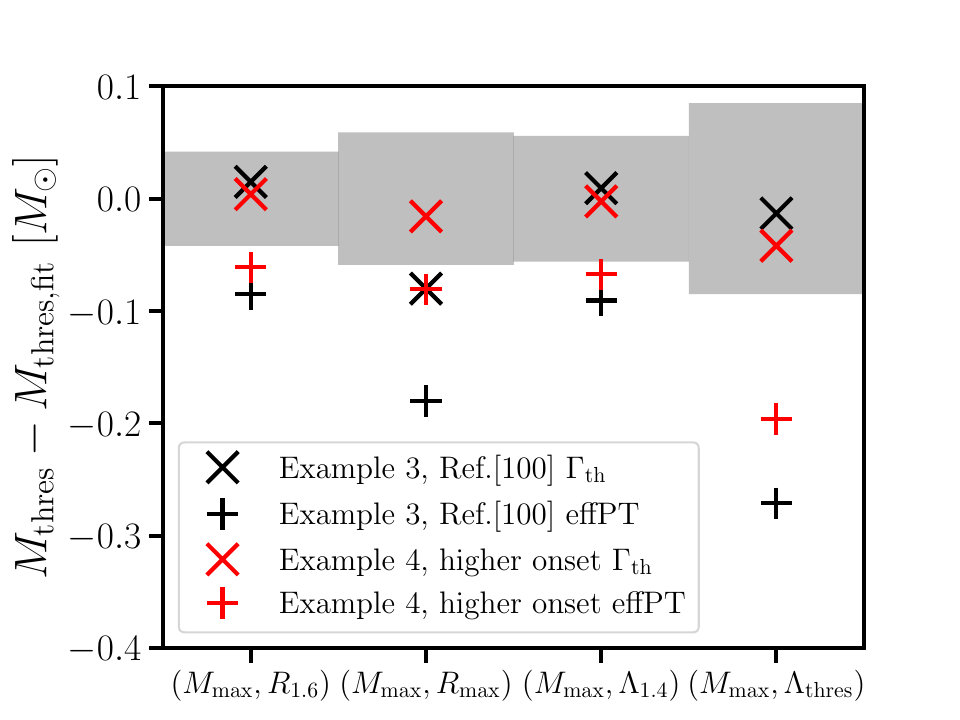}
\caption{Deviations of $M_\mathrm{thres}$ from different bilinear fits $M_\mathrm{thres,fit}(X,Y)$ with $X$ being $M_\mathrm{max}$ and $Y$ being either $R_{1.6}$, $R_\mathrm{max}$, $\Lambda_{1.4}$ or $\Lambda_\mathrm{thres}$. Fits are taken from~\cite{Bauswein2021} with the gray band indicating the respective maximum residual of the fit for purely hadronic EoS models (see~\cite{Bauswein2021} for details). The respective pair of independent variables $(X,Y)$ is given on the x-axis. Black symbols display the differences for the EoS described in Sect.~\ref{sec:otherEoSs1} (Example 3), red symbols refer to the EoS discussed in Sect.~\ref{sec:otherEoSs2} (Example 4). Crosses indicate the calculations with the $\Gamma_\mathrm{th}$ approach. Results with the effPT scheme, i.e.~with temperature-dependent phase boundaries, are displayed by plus signs.}
\label{fig:FujimotoMthresRel}
\end{figure}

Additionally, we determine the threshold mass for prompt black-hole formation for these EoSs and obtain $M_\mathrm{thres}=3.07~M_\odot$ for the $\Gamma_\mathrm{th}$ case and $M_\mathrm{thres}=3.01~M_\odot$ within the effPT treatment, which is very similar to Example 3. We compare the inferred threshold masses for Examples 3 and 4 employing both thermal treatments with the bilinear fits of Ref.~\cite{Bauswein2021}. These fits relate $M_\mathrm{thres}$ to different quantities of nonrotating stars such as the maximum mass $M_\mathrm{max}$ and NS radii or their tidal deformability (see~\cite{Bauswein2021} for details). We consider fits obtained from a set of purely hadronic EoSs (set `b' in~\cite{Bauswein2021}) to check for potential deviations that could indicate the onset of quark deconfinement.

In Fig.~\ref{fig:FujimotoMthresRel} we plot the differences we find between $M_\mathrm{thres}$ and the value predicted by the fits, $M_\mathrm{thres,fit}$, for the four different bilinear relations provided by Ref.~\cite{Bauswein2021}. The gray-shaded bands depict the maximum residuals of the fits provided by~\cite{Bauswein2021}, which thus quantify the range in which a purely hadronic model is expected to lie. The simulations with temperature-dependent phase boundaries (plus signs) lead to significantly lower $M_\mathrm{thres}$ compared to what one may expect for purely hadronic EoSs. Hence, the determination of the threshold mass may reveal the occurrence of quark matter even in such rather extreme cases like Examples 3 and 4, where signatures of quark matter are impossible or very difficult to detect in cold nonrotating NSs (see Fig.~\ref{fig:FujimotoEoSs}). The calculations with $\Gamma_\mathrm{th}$ (crosses in Fig.~\ref{fig:FujimotoMthresRel}) mostly do not yield significant deviations form the purely hadronic relations describing $M_\mathrm{thres}$, which is understandable since these systems do not reach conditions for quark deconfinement (assuming phase boundaries which do not depend on the temperature).

In general, considering postmerger features like the threshold mass or postmerger GW emission enlarges the parameter range where quark matter is astrophysically detectable. In this regard, we comment that one may expect very similar effects in a slightly less extreme scenario compared to those discussed in Example 3 and Example 4. For instance, the onset of quark deconfinement may still be hard to detect in cold NSs even for cases with stable hybrid star branch if the onset density is generally high but below the maximum density of cold nonrotating NSs. In this case only the most massive NSs will contain a quark matter core. These stars are expected not to occur very frequently in addition to the challenges to detect signatures of quark matter in such systems. As in Examples 3 and 4 one can anticipate that the temperature dependence of the phase boundaries will have a very similar impact in NS merger remnants and similarly affect the postmerger GW emission.

\section{Summary and conclusions}\label{sec:sum}

In this study we consider thermal effects of quark matter in NS mergers. The transition from hadronic matter to deconfined quark matter at finite temperature and chemical potential is generally assumed to be temperature dependent. In comparison to purely baryonic EoSs this implies a higher complexity because not only a thermal pressure component has to be modeled but also the temperature dependence of the phase boundaries has to be described. In particular, the latter can have a strong impact leading to qualitatively different results. This is because hybrid models typically exhibit a sudden change of the EoS at the phase boundary and hence the temperature dependence of the phase transition can have a very significant influence. 

A thorough investigation of these aspects is currently not straightforward because only a limited number of hybrid EoS models is available which consistently include temperature effects. There exists an approximate treatment of thermal effects in NS merger simulations, which includes an ideal-gas component to describe thermal pressure. 
This scheme is employed to supplement barotropic EoSs at $T=0$, where a large variety of models exist. However, while this treatment is successfully used for baryonic EoS models, its applicability to hybrid EoSs is questionable because it does not model the temperature dependence of the phase boundaries and thus cannot capture the major effects of the thermal properties of hybrid models. In fact, we show that this scheme does not qualitatively reproduce the behavior of temperature-dependent hybrid EoSs because it cannot correctly describe the significant softening of the EoS at finite temperature. Specifically, hybrid models typically feature an `earlier' onset of quark deconfinement at finite temperature, i.e.~at a smaller density as compared to $T=0$, which effectively leads to a substantial reduction of the pressure.

In order to explore thermal effects in NS merger simulations with hybrid EoS models, we describe here an extension of the approximate treatment of the thermal behavior which is applicable to barotropic $T=0$ hybrid EoS models. Our new effective phase transition scheme relies on a quantitative description of the phase boundaries at finite temperature, which has to be provided independently. By this it is possible to correctly describe the behavior of the coexisting phases at nonzero temperature, i.e.~the transition region between purely hadronic and pure quark matter. The procedure also allows to adopt a different effective thermal ideal gas index in the quark phase, which is usually lower than that of purely hadronic matter and closer to 4/3. This yields yet another improvement of the description of thermal effects in hybrid models compared to the traditional approach. We explicitly assume that the phase transition between the EoS of purely hadronic matter and pure deconfined quark matter is of first order and described by a Maxwell construction, i.e.~by matching pressure and chemical potential at a fixed temperature of both phases at the phase boundaries. This is a common choice in many hybrid EoS models and leads to a region of constant pressure in the transition region. We remark that there are alternative constructions of the phase transition that only require global charge conservation~\cite{Glendenning1992,Hempel2013} and result in charged, coexisting phases. We leave the extension of the ideal-gas approach to these constructions for future work.

We assess the new effPT scheme by directly comparing to a set of temperature-dependent hybrid EoS models. For this comparison we adopt a slice of the EoS at zero temperature and neutrinoless beta-equilibrium and provide the temperature dependence of the phase boundaries of the respective hybrid model by hand. We find an improvement of about an order of magnitude in the relative errors of the estimated pressure compared to the traditional ideal-gas approach.

We have further validated our effPT scheme by performing several NS merger simulations. We compare the results to simulations employing the full temperature- and composition-dependent EoS tables and to calculations with the traditional ideal-gas approach. The evolution of the merger remnant is significantly affected by the temperature dependence of the phase boundaries. Especially for the dominant postmerger GW frequency $f_\mathrm{peak}$ we find that the traditional approach potentially underestimates the frequencies by up to several hundred Hz. Generally, neglecting the temperature dependence of the phase boundaries (as in the traditional thermal ideal-gas ansatz) cannot predict the onset of quark deconfinement correctly. For instance, in some of the simulations with the simple ideal-gas treatment we observe a delayed onset of quark deconfinement after the merger and a corresponding shift in GW frequencies that are not present in simulations with the same cold EoS and an inclusion of thermal effects by either the full 3D table or our new effective thermal treatment. This cautions that employing barotropic $T=0$ hybrid EoSs in combination with the traditional thermal ideal-gas approach may not necessarily yield reliable results in NS merger simulations.

We find that our effective scheme performs well at capturing the effects of temperature-dependent phase boundaries, i.e.~an `earlier' onset of deconfinement at nonzero temperatures. Regarding the general dynamics of the merger remnant and the GW signal we see very good agreement between merger simulations using the full EoS table and simulations employing our effective scheme. As a more stringent test, we benchmark our procedure by simulating mergers with different masses and thus cover a range of regimes with no or hardly any quark deconfinement to systems with significant amounts of quark matter. Quantitatively reproducing the simulation results from the full 3D hybrid table for this whole range demonstrates that our effPT scheme correctly captures the onset of quark deconfinement, the behavior of a phase transition and quark matter at finite temperature.

These tests show that the extended scheme is suitable to evaluate thermal effects of hybrid EoSs. In particular, we investigate the impact of varying phase boundaries at finite temperature in NS mergers for a fixed barotropic EoS. Since within our effPT scheme the shape of the phase boundaries has to be provided either by analytic functions or tabulated values, we devise a simple model to describe the phase transition at finite temperature. For this we develop a parametric model for the thermal pressure, energy and chemical potential in both phases (see App.~\ref{sec:TypelModel}). Employing a Maxwell construction a specific choice of parameters then fixes the transition region at all finite temperatures.
This allows us to construct phase boundaries depending on the chosen parameters and explicitly shows that the shape of the phase boundaries at finite temperature is not fixed by the knowledge of the cold hybrid EoS alone. Hence, with this model we are able to construct very different phase boundaries at finite temperature for the same cold barotropic hybrid EoS, i.e.~with the same properties of the phase transition at zero temperature.

We demonstrate for the DD2F-SF-7 model~\cite{Bauswein2019,Bastian2021} (restricted to $T=0$) that varying the phase boundaries at finite temperature can have a substantial impact on the postmerger remnant significantly affecting the overall remnant structure and the GW signal. We find differences up to about 650~Hz in $f_\mathrm{peak}$ between simulations employing different finite-temperature phase boundaries but the same properties at $T=0$.

Adopting another example EoS from the literature based on a piecewise-polytropic model from Ref.~\cite{Weih2019}, we show that even assuming a moderate change of the phase boundaries with temperature has the potential to drastically change the merger dynamics. Specifically, we find a difference of about 1600~Hz in $f_\mathrm{peak}$ when comparing our simulation with temperature-dependent phase boundaries to a simulation employing the traditional ideal-gas approach, which neglects a temperature dependence of the phase boundary. For this model we also find that depending on the shape of the phase boundaries the occurrence of a delayed onset of quark deconfinement can happen on shorter timescales or be removed entirely as certain shapes lead to the formation of a quark core directly after the merger. 

Finally, we explicitly show that even if the hadron-quark phase transition takes place at very high densities not occurring in isolated NSs or only in the most massive stars at $T=0$, deconfined quark matter could still be present in temporarily stable NS merger remnants. This can even occur for EoS models for which no stable hybrid branch of nonrotating NSs exists. Such an outcome is possible if the boundaries of the deconfined quark matter phase shift towards lower densities at finite temperatures, which are reached in NS mergers shortly after the collision. Hence, it might be that deconfined quark matter is only accessible in finite-temperature systems such as merger remnants and proto-NSs but not in cold NSs in isolation or in binaries. Specifically, we find that in this scenario the threshold mass for prompt black-hole formation may be characteristically reduced, which would be indicative of the presence of quark matter in merger remnants. Similarly, we observe an increase of postmerger GW frequencies compared to purely hadronic systems.

All these findings demonstrate that thermal effects are more significant in hybrid than in purely hadronic EoS models as phase boundaries varying with temperature qualitatively change the behavior of the EoS. This important observation illustrates the significance of temperature effects of the QCD phase diagram in NS mergers. It exemplifies the value of observables from the postmerger phase like $f_\mathrm{peak}$ to access the finite-temperature regime of the phase diagram as a messenger providing information complementary to what can be inferred from observing cold NSs e.g.~during the late binary inspiral phase. At the same time these findings highlight that heavy-ion experiments such as HADES~\cite{Hades2019} and future facilities like FAIR~\cite{Friman2011,Senger2021} and NICA~\cite{Blaschke2016,Abgaryan2022} at finite temperature are highly desirable to better understand the impact of quark matter in NS mergers. For instance, the exclusion of deconfinement in a certain regime can be incorporated in models such as ours and thus provide important constraints.

Our effPT scheme provides the flexibility to combine any barotropic, zero-temperature EoS with different phase boundaries at finite temperature. This can be done much easier than constructing two separate full 3D tables for hadronic and quark matter and joining them with a phase construction. Especially, in the latter case it might not be easily possible to tune model parameters in a controllable way to obtain certain features such as a specific shape of the phase boundary. Therefore, our thermal treatment may prove to be useful for systematic explorations of quark matter in NS mergers employing cold, barotropic hybrid models, where a larger set of models is available in the literature or can be readily constructed by effective models.

\begin{appendix}

\section{Thermal toy model}\label{sec:TypelModel}
We provide a simple example to demonstrate that the cold, barotropic EoS does not completely fix the EoS at $T>0$. With this we show that different treatments of finite-temperature effects can result in different phase boundaries of a two-phase EoS with a first-order phase transition while still being thermodynamically consistent.

\subsection{Simple thermal EoS model}
We consider a system in thermodynamic equilibrium with properties determined by a temperature $T$, a particle number $N$ and a volume $V$. The relevant thermodynamic potential is the free energy $F$
\begin{align}
    F=-PV+\mu N
\end{align}
with the pressure $P$ and the chemical potential $\mu$ defined as
\begin{align}
    P=-\left.\frac{\partial F}{\partial V}\right\vert_{T,N}~~~~  \mu=\left.\frac{\partial F}{\partial N}\right\vert_{T,V}~.
\end{align}
With the entropy $S$
\begin{align}
    S=-\left.\frac{\partial F}{\partial T}\right\vert_{V,N}
\end{align}
the internal energy $E$ of the system is given by
\begin{align}
    E=F+TS~.
\end{align}
It is useful to remove V and N as explicit variables and consider densities
\begin{align}
    n=\frac{N}{V},~~~~ 
    f=\frac{F}{V},~~~~ 
    e=\frac{E}{V},~~~~ 
    s=\frac{S}{V}~~~~ 
\end{align}
Then we have 
\begin{align}
    f&=-P+\mu n,~~~~ 
    P=n^2\left.\frac{\partial(f/n)}{\partial n}\right\vert_{T},~~~~
    \mu=\left.\frac{\partial f}{\partial n}\right\vert_{T}\\
    s&=-\left.\frac{\partial f}{\partial T}\right\vert_{n},~~~~
    e=f+Ts=Ts-P+\mu n
\end{align}
If we split the energy into a cold part $e_\mathrm{c}$ and a thermal part $e_\mathrm{th}$ we get
\begin{align}
    f=e_\mathrm{c}+e_\mathrm{th}-Ts=f_\mathrm{c}+f_\mathrm{th}
\end{align}
with 
\begin{align}
    f_\mathrm{c}=e_\mathrm{c},~~~~f_\mathrm{th}=e_\mathrm{th}-Ts~.
\end{align}
With this we can write $e_\mathrm{th}$ and the thermal pressure $P_\mathrm{th}$ as
\begin{align}
    P_\mathrm{th}=n\left.\frac{\partial f_\mathrm{th}}{\partial n}\right\vert_{T}-f_\mathrm{th},~~~~
    e_\mathrm{th}=f_\mathrm{th}-T\left.\frac{\partial f_\mathrm{th}}{\partial T}\right\vert_{n}
\end{align}
If we assume that the relation $P_\mathrm{th}=(\Gamma_\mathrm{th}-1)e_\mathrm{th}$ holds for the thermal contribution to an EoS we obtain
\begin{align}
    \Gamma_\mathrm{th}&=\frac{P_\mathrm{th}}{e_\mathrm{th}}+1=\frac{n \left.\frac{\partial f_\mathrm{th}}{\partial n}\right\vert_{T}-f_\mathrm{th} }{f_\mathrm{th}-T\left.\frac{\partial f_\mathrm{th}}{\partial n}\right\vert_{n}}+1\\
    &=\frac{n \left.\frac{\partial f_\mathrm{th}}{\partial n}\right\vert_{T}- T\left.\frac{\partial f_\mathrm{th}}{\partial T}\right\vert_{n}  }{f_\mathrm{th}-T\left.\frac{\partial f_\mathrm{th}}{\partial n}\right\vert_{n}}
\end{align}
This expression can be written as a partial differential equation for the thermal part of the free energy density $f_\mathrm{th}$
\begin{align}
    \Gamma_\mathrm{th}f_\mathrm{th}=n\left.\frac{\partial f_\mathrm{th}}{\partial n}\right\vert_{T}-(1-\Gamma_\mathrm{th})T\left.\frac{\partial f_\mathrm{th}}{\partial T}\right\vert_{n}.
\end{align}
A solution to this equation is given by
\begin{align}
    f_\mathrm{th}=Tn \left(a \ln \frac{n}{n_0}-b\ln \frac{T}{T_0}+c\right) \label{eq:fth}
\end{align}
with three dimensionless free parameters $a>0$, $b>0$ and $c$ and two scaling variables $n_0$ and $T_0$ to guarantee proper units in the arguments of the logarithmic functions. 

This solution is similar to the free energy density of a classical ideal gas (without the rest-mass contribution) 
\begin{align}
    f_\mathrm{id}(T,n)=Tn[\ln(n\lambda^3)-1]
\end{align}
with the thermal wavelength
\begin{align}
    \lambda=\sqrt{\frac{2\pi \hbar^{2}}{mT}}
\end{align}
depending on the mass of the particles and the temperature.

From Eq.~\eqref{eq:fth} we get
\begin{align}
    \mu_\mathrm{th}&=T\left(a\ln \frac{n}{n_0}-b\ln \frac{T}{T_0}+a+c\right) \label{eq:muthsimple}\\
    P_\mathrm{th}&=aTn,~~~~
    e_\mathrm{th}=bTn,~~~~
    \Gamma_\mathrm{th}=1+\frac{a}{b} \label{eq:restsimple}
\end{align}
A specific choice of the parameters $a>0$, $b>0$ and $c$ hence fixes the thermal contribution of the EoS.
For a classical ideal gas we have $a=1$, $b=3/2$, $\Gamma=5/3$ and
\begin{align}
    c=\ln\left(n_0 \sqrt{\frac{2\pi}{mT_0}}^3 \right)-1 \label{eq:cpardef}
\end{align}
In order to recover this ideal gas limit we can set
\begin{align}
    a=\sqrt{\frac{3}{2}(\Gamma_\mathrm{th}-1)},~~~~
    b=\sqrt{\frac{3}{2(\Gamma_\mathrm{th}-1)}} \label{eq:abscale}
\end{align}

\subsection{Phase transition}\label{subsec:pt}
We now consider a system featuring a phase transition with a region of coexisting phases between particle densities $n_h(T)$ and $n_q(T)$ at a given $T$ within our simple model. We assume $n_h(T)<n_q(T)$. For this we use a standard Maxwell construction where the pressure is constant for $n_h(T)\leq n\leq n_q(T)$.
From the Maxwell relation
\begin{align}
    \left.\frac{\partial P}{\partial n}\right\vert_{T}=n\left.\frac{\partial \mu}{\partial n}\right\vert_{T}
\end{align}
we get
\begin{align}
    \mu_{h}(T)-\mu_{q}(T)=\int_{n_h(T)}^{n_q(T)}\frac{1}{n} \left.\frac{\partial P}{\partial n}\right\vert_{T} dn~.
\end{align}
Because $P=P_\mathrm{pt}(T)$ is constant with the particle density in the coexistence region the chemical potentials $\mu_{h}(T)$ and $\mu_{q}(T)$ at the borders of this region are identical $\mu_\mathrm{pt}(T)=\mu_{h}(T)=\mu_{q}(T)$ with $\mu_{h}(T)=\mu(T,n_{q}(T))$ and $\mu_{q}(T)=\mu(T,n_{q}(T))$.

If we write the free energy densities of both phases each as the sum of a $T=0$ component and a thermal contribution
\begin{align}
    f^h=e_\mathrm{c}^h+f_\mathrm{th}^h,~~~~
    f^q=e_\mathrm{c}^q+f_\mathrm{th}^q
\end{align}
and parametrize the thermal contributions according Eq.~\eqref{eq:fth} we get
\begin{align}
\begin{split}
    P_\mathrm{pt}(T)=&P_\mathrm{c}(n_h(T))+a_h Tn_h(T)\\
    =&P_\mathrm{c}(n_q(T))+a_q Tn_q(T)\label{eq:pth1}
\end{split}\\
\begin{split}
    \mu_\mathrm{th}(T)=&\mu_\mathrm{c}(n_h(T))+T(a_h\ln(n_h/n_0)- \\ &b_h\ln(T/T_0)+a_h+c_h)\\
    =&\mu_\mathrm{c}(n_q(T))+T(a_q\ln(n_q/n_0)- \\ &b_q\ln(T/T_0)+a_q+c_q)~. \label{eq:muth1}
\end{split}
\end{align}
These two equations fix the phase boundaries $n_h(T)$ and $n_q(T)$ at a given temperature.

The energy densities at the borders of the coexistence region are given by
\begin{align}
    e(T,n_h(T))=e_{c,h}(n_h(T))+b_h Tn_h(T)\label{eq:eth1}\\
    e(T,n_q(T))=e_{c,q}(n_q(T))+b_q Tn_q(T)\label{eq:eth2}
\end{align}
where $e_{c,h}$ and $e_{c,q}$ are the energy densities of both phases at $T=0$.

The choice of the parameters $a_h$, $b_h$, $c_h$, $a_q$, $b_q$ and $c_q$ together with both phases at $T=0$ therefore completely determines the borders of the phase transition region.

Note that $\Gamma_{\mathrm{th},h}$ and $\Gamma_{\mathrm{th},q}$ are fixed by the relation $\Gamma_\mathrm{th}=1+a/b$. 

\subsection{Example phase boundaries}

From this discussion above we see that picking different parameters for $f_\mathrm{th}$ will change the behavior of the phase boundaries at $T>0$ for the same cold EoS. Therefore the EoS at $T=0$ alone does not fix the transition region at $T>0$. 

As an example we consider the following simple two-phase model. We assume a cold polytropic EoS representing the hadronic regime at lower densities and another cold polytropic EoS describing deconfined quark matter at higher densities. Both EoSs can hence be described by
\begin{align}
    p_\mathrm{c}&=Kn^{\Gamma}\\
    \mu_\mathrm{c}&=\frac{K\Gamma}{\Gamma-1}n^{\Gamma-1}+E_0~.
\end{align}
We pick $\Gamma=1.8$ for the low density phase and $\Gamma=2.5$ for the high density phase. We also require that the coexistence phase lies between $n_\mathrm{on}=0.3~\mathrm{fm}^{-3}$ and $n_\mathrm{fin}=0.5~\mathrm{fm}^{-3}$, has a pressure of $p_\mathrm{pt}=21~\mathrm{MeV/fm}^{3}$ and a chemical potential of $\mu_\mathrm{pt}=1005~\mathrm{MeV}$. This choice of $p_\mathrm{pt}$ and $\mu_\mathrm{pt}$ at $n_\mathrm{on}$ and $n_\mathrm{fin}$ fixes the parameters $K$ and $E_0$ in both cold phases.

We construct the finite-temperature part with our simple approach using Eq.~\eqref{eq:muthsimple} and Eq.~\eqref{eq:restsimple}. 
For the low density phase we set $\Gamma_\mathrm{th}=1.75$ and for the high density phase $\Gamma_\mathrm{th}=4/3$. We then determine $a_h$ and $b_h$ of the low density phase using Eq.~\eqref{eq:abscale}. For the high density phase we pick different values of $a_q$ and determine $b_q$ from the relation $\Gamma_\mathrm{th}=1+a/b$. In both phases we calculate the parameter $c$ using Eq.~\eqref{eq:cpardef} with $m=931.49$~MeV and set the scaling variables to $n_0=1~\mathrm{fm}^{-3}$ and $T_0=1~\mathrm{MeV}$.

We then construct the phase boundary for each set of parameters using the conditions from Eq.~\eqref{eq:pth1} and Eq.~\eqref{eq:muth1}. For the tabulated DD2F-SF EoSs we find that the parameter $a$ lies in the range of $0.1-0.85$ for densities between $3.5\times \rho_\mathrm{nuc}$ and $7\times \rho_\mathrm{nuc}$ and temperatures between 10~MeV and 50~MeV. Therefore, we pick the four different values of $a_\mathrm{q}$ from this range.

\begin{figure}
\centering
\includegraphics[width=1.0\linewidth]{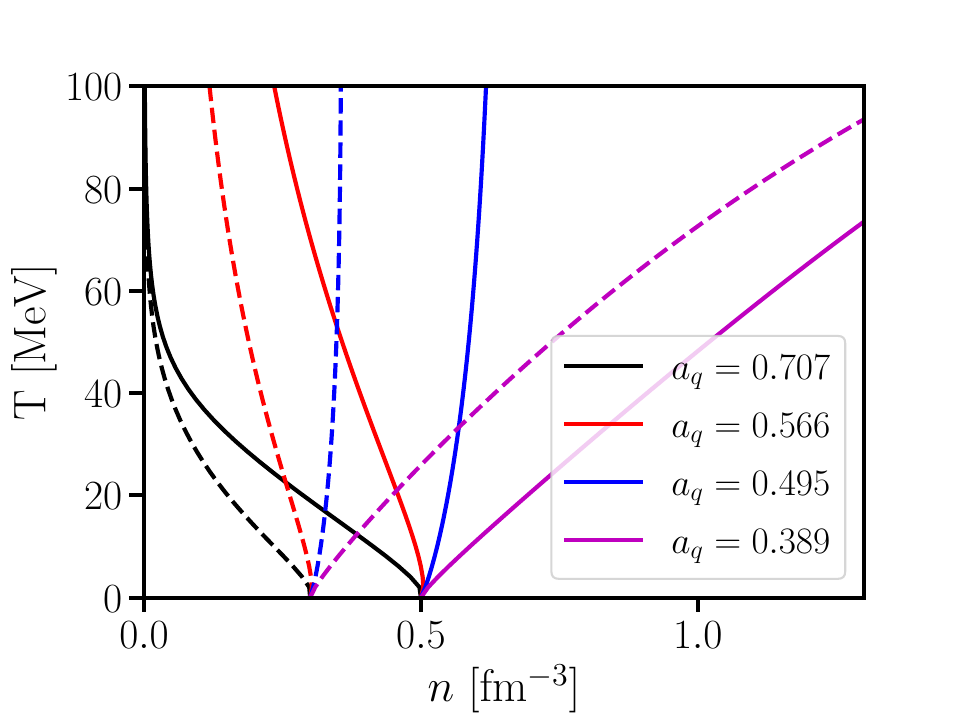}
\caption{Phase boundaries in the $n$-$T$-plane determined with our simple phase transition model for different choices of $\Gamma_\mathrm{th}$ in the high density phase for a single cold EoS. Different colors refer to different phase boundaries. Dashed lines mark the beginning of the coexisting phases and solid lines display the onset of pure deconfined quark matter.}
\label{fig:toybounds}
\end{figure}

In Fig.~\ref{fig:toybounds} we show the phase boundaries we obtain with this model for the different values of $a_q$ in the $n-T$ diagram with different colors. We plot the onset densities of the phase transition with dashed lines and the beginning of the high density phase with solid lines.

We find that within this simple model we can generate a large variety of phase boundaries with qualitatively different behaviors at finite temperature. For example for $a_q=0.707$ we see that the boundaries of the coexistence phase strongly shift towards lower densities for finite temperature. On the other hand picking $a_q=0.389$ moves the phase boundaries to larger densities with increasing $T$. Note that at $T=0$ all examples have the same $\rho_\mathrm{on}$ and $\rho_\mathrm{fin}$ as the cold EoS is identical in every case.

This finding demonstrates that the EoS at $T=0$ does not uniquely determine the behavior of the phase boundaries at $T>0$. Different assumptions on finite-temperature effects can lead to qualitatively different behaviors of the transition region.

We also stress again that all examples have $\Gamma_\mathrm{th}=1.75$ in the low density phase and $\Gamma_\mathrm{th}=4/3$ in the high density phase. Hence, a specific choice of $\Gamma_\mathrm{th}$ in both phases does also not fix the phase boundaries at finite temperature. 

\section{Shifted phase boundaries for the DD2F-SF-7 EoS}\label{sec:shiftedbounds}
Here we provide the construction of the different phase boundaries for the DD2F-SF-7 EoS at finite temperature we used in Sect.~\ref{sec:impact}. At $T=0$ this model has coexisting phases between $\rho_\mathrm{on,0}=3.5\times \rho_\mathrm{nuc}$ and $\rho_\mathrm{fin,0}=3.7\times \rho_\mathrm{nuc}$.

We use our approach of App.~\ref{sec:TypelModel} to model the thermal part of the EoS where we pick different parameters for the hadronic and the quark phase. Additionally, we need to describe the cold EoS of both phases at all densities to balance the total pressure and chemical potential for a given temperature.

We find that in the density range relevant for constructing the phase transition between $0.1\times \rho_\mathrm{nuc}$ and $3.5\times \rho_\mathrm{nuc}$ the DD2F EoS can be approximated well by a piecewise polytropic EoS with two segments. In such a model the pressure, chemical potential and energy density of each segment are given by
\begin{align}
    P&=K_{i}n^{\Gamma_{i}} \label{eq:Ppp}\\
    \mu&=\frac{K_{i}\Gamma_{i}}{\Gamma_{i}-1}n^{\Gamma_{i}-1}+E_{0,i}\label{eq:mupp}\\
    e&=\frac{K_{i}}{\Gamma_{i}-1}n^{\Gamma_{i}}+E_{0,i}n\label{eq:epp}
\end{align}
with $i=1,2$. We perform a least squares fit to the tabulated pressure of the DD2F EoS with Eq.~\eqref{eq:Ppp} in the aforementioned density range. During this fit we pick the parameters $K_{2}$ and $E_{0,2}$ to recover $P$ and $\mu$ of the DD2F EoS at $n=3.5\times \rho_\mathrm{nuc}$. This ensures that we recreate the true onset density of the DD2F-SF-7 EoS at $T=0$. The parameters $K_{1}$ and $E_{0,1}$ are fixed by requiring continuous $P$ and $e$. Hence we are left with fitting the parameters $\Gamma_{1,2}$ and the density $n_1$ where we switch from one segment to the other.

For the cold quark phase a piecewise polytropic model is not well suited because the pressure is expected to become negative at low densities due to quark confinement. Such a behavior cannot be achieved in a polytropic approach. Instead we describe the quark phase using the bag model of~\cite{Alford2005}. In this model the grandcanonical potential density $\Omega$ is given by
\begin{align}
    \Omega=-\left(\frac{3}{4\pi^{2}}a_4\mu_q^4+\frac{3}{4\pi^{2}}a_2\mu_q^2+B_\mathrm{eff}\right)\frac{1}{(\hbar c)^{3}}~.
\end{align}

From this we obtain
\begin{align}
    P_q&=\left(\frac{3}{4\pi^{2}}a_4\mu_q^4-\frac{3}{4\pi^{2}}a_2\mu_q^2-B_\mathrm{eff}\label{eq:pbag}\right)\frac{1}{(\hbar c)^{3}}\\
    n_q&=\left(\frac{3}{\pi^{2}}a_4\mu_q^3-\frac{3}{2\pi^2}a_2\mu_q\right)\frac{1}{(\hbar c)^{3}}  \label{eq:nbag}\\
    e_q&=\left(\frac{9}{4\pi^2}a_4\mu_q^4-\frac{3}{4\pi^2}a_2\mu_q^2+B_\mathrm{eff}\right)\frac{1}{(\hbar c)^{3}}~. \label{eq:ebag}
\end{align}
Note that here $\mu_q$ and $n_q$ refer to the quark chemical potential and number density, respectively. We assume $3\mu_q=\mu$ and $n_q=3n$.

We obtain the parameter $a_4$ with a least squares fit using Eq.~\eqref{eq:pbag} and the tabulated pressures of the DD2F-SF-7 EoS in the density range between $\rho_\mathrm{on,0}=3.7\times \rho_\mathrm{nuc}$ and $\rho_\mathrm{fin,0}=15.0\times \rho_\mathrm{nuc}$. In order to recover the correct phase boundary at $T=0$, i.e.~to exactly match $P$ and $\mu$ of the DD2F-SF-7 EoS at this density we determine the parameters $a_2$ and $B_\mathrm{eff}$ during the fitting from the tabulated values of $P$ and $\mu$ at $n=3.7\times n_\mathrm{nuc}$ and the current value of $a_4$. We provide the parameters for the cold hadronic phase in Tab.~\ref{tab:fitparsH} and the parameters we use for the cold quark phase in Tab.~\ref{tab:fitparsQ}.

\begin{table}
\begin{tabular}{c c c c c c c}
\hline\hline
$K_1$ & $K_2$ & $\Gamma_2$ & $\Gamma_2$ & $E_{0,1}$ & $E_{0,2}$ & $n_1$\\
$\left[ \mathrm{MeV}\right] $ & $\left[ \mathrm{MeV}\right] $ & & & $\left[ \mathrm{MeV}\right] $ & $\left[ \mathrm{MeV}\right] $ & $\left[ \mathrm{fm}^{-3}\right] $\\
\hline\
30.88 & 561.9 & 1.727 & 2.832 & 940.0 & 943.8 & 0.072  \\
\hline
\hline
\end{tabular}
\caption{Parameters we use to describe the hadronic phase with a two segment piecewise polytropic EoS (see Eqs.~\eqref{eq:Ppp}-\eqref{eq:epp}).}
\label{tab:fitparsH}
\end{table}

\begin{table}
\begin{tabular}{c c c}
\hline\hline
$a_4$ & $a_2$ & $B_\mathrm{eff}$ \\
$ $ & $ \left[ \mathrm{MeV}^{2}\right] $ & $ \left[ \mathrm{MeV}^{4}\right] $\\
\hline\
$0.117$ & $-(419.0)^2 $ & $203.8^4$ \\
\hline
\hline
\end{tabular}
\caption{Parameters we use to describe the quark phase with a bag model EoS (see Eqs.~\eqref{eq:pbag}-\eqref{eq:ebag}).}
\label{tab:fitparsQ}
\end{table}

To find the phase boundaries at finite temperature we pick $\Gamma_\mathrm{th}=1.75$ for the hadronic phase and $\Gamma_\mathrm{th}=4/3$ for the quark phase. Using $\Gamma_\mathrm{th}=1+a/b$ this fixes the parameter $b$ for a choice of $a$. 

To construct the phase boundaries of model low in Sect.~\ref{sec:impact} we pick $a_h=1.060$ and $c_h=8.456$ for the hadronic phase and $a_q=0.707$ and $c_q=6.256$ for the quark phase. For model high we choose $a_h=1.167$, $c_h=7.356$, $a_q=0.658$ and $c_q=7.356$.

As a comparison, we determine the effective values of the parameters $a$ and $c$ from the tabulated pressure and chemical potential of the hadronic DD2F and the hybrid DD2F-SF EoS in the temperature interval 10~MeV to 50~MeV. For the DD2F-SF EoSs we find that the parameter $a$ lies in the range of $0.1-0.85$ for densities between $3.5\times \rho_\mathrm{nuc}$ and $7\times \rho_\mathrm{nuc}$. The parameter $c$ can vary between around $0.9$ and $11$. For the hadronic DD2F EoS we observe values of $a$ from $0.1$ to $1.5$ and values of $c$ from $0.4$ to $20$ in the density range from $0.1\times \rho_\mathrm{nuc}$ to $3.5\times \rho_\mathrm{nuc}$. The parameters we have chosen to construct the shifted phase boundaries are hence within the range of values we observe for the tabulated EoS.

With these descriptions of the cold and the thermal parts of our two phases we construct the phase boundaries using Eq.~\eqref{eq:pth1} and Eq.~\eqref{eq:muth1}. Because we need the phase boundaries in the $\epsilon_\mathrm{th}$-$n$ plane we calculate the energy densities at the phase boundaries using Eq.~\eqref{eq:eth1} and Eq.~\eqref{eq:eth2}.

We plot the phase boundaries of both model low and model high in Fig.~\ref{fig:NewBounds484} in the $\epsilon_\mathrm{th}-n$ plane with blue and red lines, respectively. For a comparison we also plot the true boundaries of the DD2F-SF-7 EoS, i.e.~model true, with black lines. Dashed (solid) lines mark the onset (end) of the coexistence phase.

We find that for model low our choice of parameters shifts the coexistence phase to lower densities compared to model true. On the other hand we see that in model high the onset of quark deconfinement moves to larger densities while the end of the coexistence phase is almost identical compared to model true.

\end{appendix}

\acknowledgements{We thank T. Fischer and D. A. Terrero for helpful discussions. This work was funded by Deutsche Forschungsgemeinschaft (DFG, German Research Foundation) - Project-ID 279384907 - SFB 1245. AB acknowledges support by the European Research Council under the European Union’s Horizon 2020 research and innovation programme under Grant No. 759253 and by the State of Hesse within the Cluster Project ELEMENTS.}

\bibliography{references}

\end{document}

%% file: journalheader.tex
\newcommand{\apjl}{Astrophys. J. Lett.}%
\newcommand{\aap}{Astron. Astrophys.}%
\newcommand{\mnras}{Mon. Not. Roy. Astron. Soc.}%
\newcommand{\pasa}{Publications of the Astronomical Society of Australia}
\newcommand{\physrep}{Phys. Rept.}
\newcommand{\nphysa}{Nucl. Phys. A}